\documentclass[11pt,a4paper]{article}              

\usepackage[margin=25mm]{geometry}
\usepackage{graphicx}
\usepackage{colortbl}
\usepackage{fancyhdr}
\usepackage{amssymb,amsfonts,amsmath,amsthm}
\usepackage{natbib}

\usepackage{authblk}
	
\usepackage{subfig}
\usepackage{multirow}
\usepackage{amssymb,amsfonts,amsmath,amsthm}
\usepackage{bbm}
\graphicspath{{./},{./Figures/}}
\usepackage[noend]{algpseudocode}
\usepackage{algorithm}
\usepackage{algorithmicx}
\algrenewcommand\algorithmicrequire{\textbf{Initialization:}}
\usepackage{xcolor}
\definecolor{DarkGreen}{RGB}{0,100,0}   
\definecolor{DarkBlue}{RGB}{65,105,225} 

\newcommand{\ve}[1]{\boldsymbol{#1}}			

\newcommand{\enum}{ , \, \dots \,,}

\newcommand{\prt}[1]{\left(#1\right)}			
\newcommand{\acc}[1]{\left\{#1\right\}}			
\newcommand{\bra}[1]{\left[ #1 \right]}			
\newcommand{\abs}[1]{\left| #1 \right|}			

\newcommand{\ie}{{\em i.e.} }

\begin{document}
\title{Reliability analysis of arbitrary systems based on active learning and global sensitivity analysis} 

\author[1]{M. Moustapha} \author[1]{P. Parisi} \author[1]{S. Marelli}  \author[1]{B. Sudret}   

\affil[1]{Chair of Risk, Safety and Uncertainty Quantification,
  
  ETH Zurich, Stefano-Franscini-Platz 5, 8093 Zurich, Switzerland}

\date{}
\maketitle

\abstract{System reliability analysis aims at computing the probability of failure of an engineering system given a set of uncertain inputs and limit state functions. Active-learning solution schemes have been shown to be a viable tool but as of yet they are not as efficient as in the context of component reliability analysis. This is due to some peculiarities of system problems, such as the presence of multiple failure modes and their uneven contribution to failure, or the dependence on the system configuration (e.g., series or parallel). 
In this work, we propose a novel active learning strategy designed for solving general system reliability problems. 
This algorithm combines subset simulation and Kriging/PC-Kriging, and relies on an enrichment scheme tailored to specifically address the weaknesses of this class of methods. More specifically, it relies on three components: (i) a new learning function that does not require the specification of the system configuration, (ii) a density-based clustering technique that allows one to automatically detect the different failure modes, and (iii) sensitivity analysis to estimate the contribution of each limit state to system failure so as to select only the most relevant ones for enrichment. The proposed method is validated on two analytical examples and compared against results gathered in the literature. Finally, a complex engineering problem related to power transmission is solved, thereby showcasing the efficiency of the proposed method in a real-case scenario. \\[1em] 

}

\maketitle

\section{Introduction}
Engineering systems generally operate in an uncertain environment. Quantifying the impact of these uncertainties is paramount to designing safe structures. 
Reliability analysis aims precisely at predicting the probability of failure of a structure, in which failure is characterized as the violation of some pre-defined limit state function. The latter maps the set of uncertain parameters describing the structure (\emph{e.g.}, material properties) and its environment (\emph{e.g.}, loads) to a scalar value whose sign encodes the state of the system, i.e., whether it is in failure or safety state.

Computing the failure probability is ultimately equivalent to solving a multi-dimensional integration problem, which is generally not tractable analytically. It is instead solved using approximation techniques such as the first- and second-order reliability methods or Monte Carlo simulation \citep{Ditlevsen1996,Lemaire2009,Melchers2018}. 
While approximation techniques are relatively inexpensive, they lack accuracy in cases that involve high-dimensional or non-linear limit states. 
In contrast, simulation methods can yield accurate results and provide confidence bounds on their estimators, but are computationally intensive, and hence not suitable when the limit state function relies on expensive computational models. 
In the last decade, they have been complemented by surrogate models, which act as inexpensive approximations of the limit state function. Used in the so-called \emph{active learning} framework, surrogate models have been shown to efficiently solve reliability analysis problems \citep{Teixeira2021,MoustaphaSS2022}. Examples include active-Kriging Monte Carlo simulation (AK-MCS, \citet{Echard2011}) or efficient global reliability analysis (EGRA, \citet{Bichon2009}), which both use Kriging a.k.a. Gaussian process model \citep{Santner2003, Rasmussen2006} as a surrogate.

In active learning, a surrogate model is trained using a limited set of evaluations of the original model, known as the \emph{experimental design}. The goal is to ensure its accuracy in the vicinity of the limit state surface (i.e., the boundary between the failure and safety domains) in regions of high probability density. To this end, a \emph{learning function} is used to adaptively and parsimoniously select sets of input parameters that, when added to the experimental design, are most likely to increase the accuracy of the limit state surface approximation and, henceforth, of the estimated probability of failure. 

This is an active field of research as shown by the recent profusion of methods in the literature (for a comprehensive review and benchmark analysis, see \citet{Teixeira2021} and \citet{MoustaphaSS2022}, respectively). 
These methods are mainly developed for so-called \emph{component} reliability problems, where failure is described by a single limit state function. Yet in complex engineering systems, multiple limit state functions may be required to properly capture different failure modes. This problem is known as \emph{system} reliability analysis. 
In contrast to component reliability, there are fewer contributions in the literature, most of which do not address the specific issues related to system reliability, such as the presence of multiple disjoint failure domains or their uneven contribution to system failure.

The most straightforward way to tackle system reliability problems using active learning is to directly
approximate the system performance function, which is a combination of the component limit states.  By doing
so, one solves again a component reliability problem while reducing the number of models to approximate and
evaluate to a single one.  However, by construction, the system limit state is a non-smooth function that
may exhibit discontinuities or high irregularities, and is thereby difficult to approximate accurately
\citep{Tresp2000,Meeds2005,Boroson2017,MoustaphaCS2023}. This may happen for instance when the system limit
state is obtained as the minimum of all the component ones (in case of so-called \emph{parallel systems}
limit states).

It is then more convenient and efficient to separately approximate each of the components. This is even more
relevant when latter rely on fundamentally different computational models (e.g., in multiphysics
analyses). Many approaches proposed in the literature do so by simply adapting component reliability
learning functions to solve system problems. For instance, AK-SYS \citep{Fauriat2014} is an extension of
AK-MCS where different ways of computing the learning function (deviation number $U$) are
proposed. \citet{Guo2021} propose an improvement of AK-SYS by considering kernel density based adaptive
importance sampling instead of MCS. AK-SYSi \citep{Yun2019} is a direct improvement of AK-SYS where the
issue of inaccurately estimating the modes of the different Kriging models is addressed.  More recently, new
learning functions have been proposed to address the system limit-state while using separate surrogates at
the component level. Examples include the expected system improvement suggested by \citet{Yang2022}, a
combination of the expected improvement and the U learning function \citep{Wang2023} or the APe
(approximation of the probability of error), which is based on the probability of misclassification
\citep{Xu2023}. Similarly, \citet{Feng2023} proposed a learning function based on the upper bound of the
probability of misclassification, taking into account the correlation between different failure modes.

Arguing that such methods fail to properly discard unimportant limit state functions when their responses
differ in magnitude, \citet{YangX2018} propose to define truncated candidate regions (TCR) where unimportant
areas are first identified. Only samples not belonging to the TCR are selected as candidate points.  The
traditional composite criterion of AK-SYS is then used to select the enrichment points and the corresponding
limit state function to evaluate. \citet{YangX2020} propose an extension of ALK-TCR considering multi-modal
adaptive importance sampling. \citet{Zhou2022} consider ALK-TCR but break down the enrichment scheme in two
different strategies according to the system configuration (series or parallel).

From another perspective, \citet{Wu2020} extend the concept of dependent Kriging models where the correlation of the component responses between different candidate points is taken into account \citep{Hu2016} while devising a new learning function for system problems (DKM-SYS). 
\citet{Wang2021} similarly propose a method relying on the misclassification at the system level. 
Accounting for the dependencies between different responses at the same location (assuming a single model run returns multiple outputs), \citet{Sadoughi2018} use multivariate Gaussian process models, while \citet{Hu2017} propose building composite Kriging models combining both the individual models and others based on an SVD decomposition of the system responses. 
Other contributions involving multivariate Gaussian processes include \citet{Perrin2016} where nested GP models are considered or \citet{Wei2018}. 

All these methods are only applicable to systems that are either parallel or series.  In practice, it is always possible to use cut-sets to transform any arbitrary configuration into a series or parallel system. Nevertheless, the reconfigured cut-sets still result in non-smooth, and henceforth difficult to approximate, limit states. More general methods capable of handling complex system configurations have been developed recently. 
For instance, \citet{Yuan2020} propose an approach where the learning function is built on a system level using system information defined through minimal path sets. 
The latter is also used by \citet{Xiao2022} to extend DKM-SYS to arbitrary system configurations. 

The aforementioned contributions mostly tackle only one or two aspects of the issues raised by system reliability. 
In this paper, we propose a new general and efficient algorithm for surrogate-based system reliability analysis that address them \emph{all at once}. 
More precisely, we devise an enrichment scheme that identifies relevant failure modes through subset simulation and density-based clustering. 
Furthermore, we use Sobol' sensitivity analysis to identify the limit state functions that contribute the most to the system failure, so as to only enrich the corresponding experimental designs. The approach can be used for general systems, \emph{i.e.}, it is not limited to series of parallel ones.

The paper is organized as follows. In Section~\ref{sec:Statement}, we state the problem, introduce useful notations and discuss the motivations for this work. 
In Section~\ref{sec:Method}, we present the proposed method and detail all its components. 
Finally in Section~\ref{sec:Examples}, we validate the proposed method using two benchmark problems and apply it to an engineering problem which consists of a system of transmission towers.

\section{Problem statement and motivations}\label{sec:Statement}

\subsection{Problem statement}
Let us consider a set of $m$ quantities of interest $Y_j$ describing the performance of a system through a set of \emph{limit state functions} $g_j$ such that
\begin{equation}\label{eq:YMX}
    Y_j = g_j\prt{\ve{X}_j}, \quad j = 1, \ldots, m,
\end{equation}
where $\ve{X}_j \in \mathcal{D}_{\ve{X}_j} \subset \mathbb{R}^{M_j}$ is an $M_j$-dimensional random vector associated to the $j$-th \emph{component} of the system.
These random variables characterize the uncertainties associated to the components of interest, such as the stochastic variability of the loads or manufacturing tolerances. 
In component reliability analysis, we are interested in finding the probability of failure of each component, which can be computed as
\begin{equation}\label{eq:Pf_comp}
P_{f_j} = \int_{\mathcal{D}_{f_j}} f_{\ve{X}_j}\prt{\ve{x}_j}\, \mathrm{d}\ve{x}_j,
\end{equation}
where $f_{\ve{X}_j}$ is the joint density of the random variables gathered in $\ve{X}_j$ and $\mathcal{D}_{f_j}$ denotes the failure domain of the $j$-th component. The latter is by convention implicitly defined through the sign of the corresponding limit state function, \emph{i.e.}, $\mathcal{D}_{f_j} = \acc{\ve{x}_j \in \mathbb{R}^{M_j}: g_j\prt{\ve{x}_j} \leq 0}$.

In this work, we are primarily interested in \emph{system reliability analysis}, where the performance of the system is jointly described by the component limit states. 
In this context, the definition of the failure domain is not straightforward and depends on the configuration of the system. 
Notable configurations include parallel and series systems. 
A parallel system is one for which system failure occurs only when all components fail, while a series system is one that fails when at least one component fails.
Assuming that the system state can be sufficiently described by the knowledge of the state of each component, any arbitrary configuration can be defined as a parallel (resp. series) arrangement of a series (resp. parallel) system using the concepts of minimal path sets (resp. minimal cut sets) \citep{Ross2010}. 

We consider here the most general case when the system configuration is defined through an arbitrary function $h\prt{g_1\prt{\ve{X}_1}, \ldots, g_m\prt{\ve{X}_m}} = h\prt{\ve{g}\prt{\ve{X}}}$. 
We introduce here the $M$-dimensional random vector describing the entire system $\ve{X} \in \mathbb{R}^{M} \sim f_{\ve{X}}(\ve{x})$ that contains all components of the vectors $\ve{X}_j$ without duplication. The support of $\ve{X}$ is obtained by the union: $\mathcal{D}_{\ve{X}} = \cup_{j=1}^{m} \mathcal{D}_{\ve{X}_j}$. This means that the random variables $\ve{X}_j$ are not necessarily the same, even though there is in general some overlap. While a Boolean composition of $\min$ and $\max$ operators may typically define the \emph{composition function} $h$, this work is not restricted to such compositions. We however assume that $h$ is an easy-to-evaluate, analytical function, more precisely that its evaluation cost is comparatively much smaller than that of any of the component limit state functions. This allows us to use sampling-based approaches in the methodology we propose

The system failure probability therefore reads
\begin{equation}\label{eq:Pf_sys}
P_{f} = \mathbb{P}\bra{h\prt{\ve{g}\prt{\ve{X}}} \leq 0} = \int_{\mathcal{D}_{f}} f_{\ve{X}}\prt{\ve{x}}\, \mathrm{d}\ve{x},
\end{equation}
where $f_{\ve{X}}$ is the joint density of $\ve{X}$ and the system failure domain $\mathcal{D}_{f}$ is defined by $\mathcal{D}_{f} = \acc{\ve{x} \in \mathbb{R}^M: h\prt{\ve{g}\prt{\ve{x}}} \leq 0}$.

\subsection{Motivation}
Active learning reliability methods are arguably the most efficient way to solve the integration problems in  Eqs.~(\ref{eq:Pf_comp})~and~(\ref{eq:Pf_sys}). 
Such methods have been thoroughly explored in the context of component reliability, as documented in recent surveys and benchmarks \citep{Teixeira2021,MoustaphaSS2022}. 
These methods have also been extended or adapted to system problems. Yet the proposed extensions are generally poor in many aspects: (i) some attempt to approximate $h\prt{\ve{g}\prt{\ve{x}}}$, which is generally a non-homogeneous or discontinuous function, (ii) others lack a consistent way of selecting the most appropriate component limit state function to enrich, resulting in enrichment points located in areas with little to no contribution to system failure, and (iii) most methods are specifically tailored to a particular system configuration (usually, series or parallel).

In this paper, we propose a methodology that aims at bypassing all these limitations. 
More precisely, the proposed method allows solving complex system reliability problems and its implementation does not depend on the system configuration or any knowledge thereof. 
Furthermore, the methodology allows for a large computational cost saving by enriching only the experimental designs corresponding to relevant limit states, \emph{i.e.}, areas that contribute to component but not to system failure are not enriched. Finally, we aim at solving problems with relatively small failure probabilities.

More precisely, we tackle the three limitations respectively by (i) building a separate surrogate model for each component, (ii) using Sobol' sensitivity analysis on $h(\widehat{\ve{g}}(\ve{x}))$ to identify the specific limit state function(s) that need to be updated, hence limiting enrichment to areas effectively contributing to system failure, and (iii) introducing a new system learning function for arbitrary system configurations. Furthermore, to estimate small failure probabilities, we use subset simulation \citep{Au2001}, which also provides us with the candidate samples for enrichment (following the recommendation of the comprehensive benchmark by \citet{MoustaphaSS2022}).


\section{Proposed method for system reliability}\label{sec:Method}

\subsection{Flowchart of the proposed method}
The proposed method draws upon the traditional active learning reliability framework for component problems, which comprises four ingredients, namely a surrogate model, a reliability estimation algorithm, an enrichment scheme and a stopping criterion \citep{MoustaphaSS2022}. These ingredients are generally assembled in a sequential way in a framework popularized by AK-MCS \citep{Echard2011}.

\begin{figure}[!ht]
    \centering\includegraphics[width=0.8\textwidth]{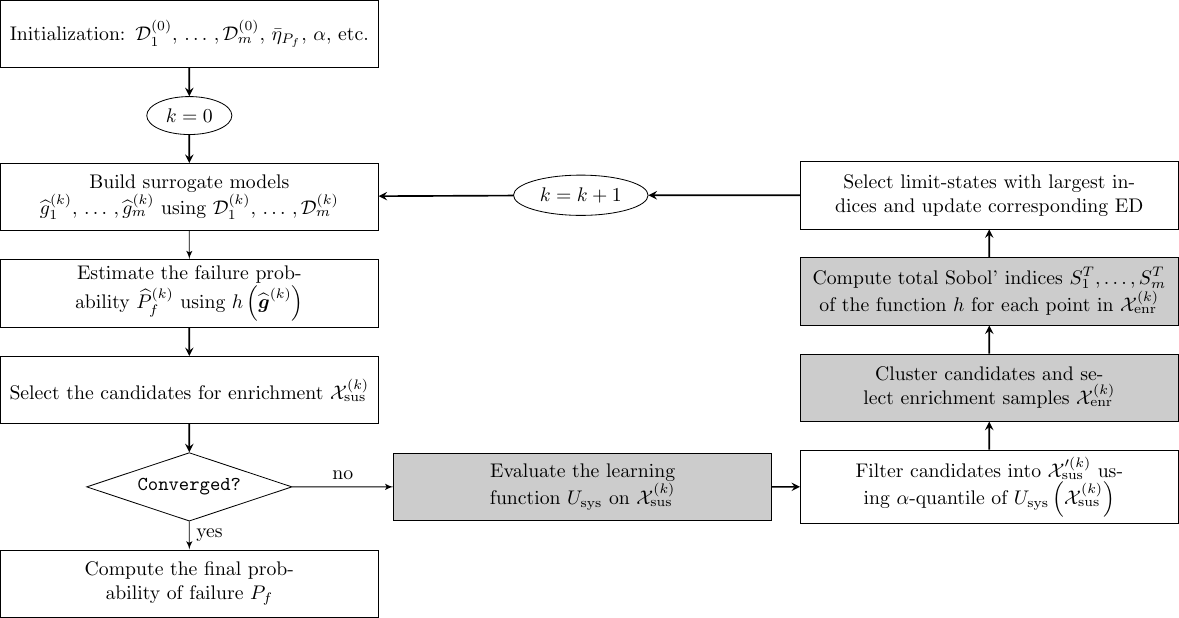}
	\caption{Flowchart of the proposed active learning method for system reliability.}
	\label{fig:SYSALR}
\end{figure}

The same framework is considered here for system reliability, as illustrated in the flowchart in Figure~\ref{fig:SYSALR}. Additional steps are however taken to address the peculiarities of system reliability. They are highlighted in gray in the flowchart.

The algorithm is first summarized in the following steps:

\begin{enumerate}
    \item \textbf{Initialization:} The initial experimental design is built by sampling the input space using a space-filling technique such as Latin hypercube sampling or Sobol' sequences \citep{McKay1979,Sobol1967}. 
    To cover as uniformly as possible the input space and reach areas far in the tails of the distribution (which are likely to be in the failure domain), we sample in a hypercube $\prod_{i=1}^{M} \bra{x_i^{-}, \, x_i^{+}}$. There are many ways to define these bounds. For low-dimensional cases, we simply consider  $x_i^{-} = \mu_{x_i} - 5 \, \sigma_{X_i}$ and $x_i^{+} = \mu_{X_i}+5\,\sigma_{X_i}$, where $\mu_{X_i}$ and $\sigma_{X_i}$ are respectively the mean and standard deviation of the $i$-th random variable. For high-dimensional cases, we consider upper and lower quantiles of the distribution in each dimension. Once the inputs are sampled, the experimental design is completed by evaluating the corresponding limit state functions on the generated samples. 
    For each limit state $j$, the experimental design is denoted by $\mathcal{D}_j = \acc{\mathcal{X}_j, \mathcal{G}_j} =  \acc{\prt{\ve{x}_j^{(i)},\, g_j \prt{\ve{x}_j^{(i)}}}, i = 1, \ldots N_{j}^{(0)}}$, for $j = 1, \ldots, m$.
    \item \textbf{Surrogates construction:} $m$ independent surrogate models $\widehat{g}_j$ are then built using the respective experimental designs $\mathcal{D}_j, \, j = \acc{1, \ldots, m}$. We consider here surrogates with built-in error measures, namely Kriging and polynomial-chaos Kriging, which are further described in Section~\ref{sec:surrogates}. 
    \item \textbf{System reliability:} Using the approximate system limit state function $h\prt{\widehat{g}\prt{\ve{x}}} = $\\ $h \prt{\widehat{g}_1\prt{\ve{x}_1}, \ldots, \widehat{g}_m\prt{\ve{x}_m}}$, the probability of failure of the system is estimated. We consider subset simulation for this task. The samples generated throughout the algorithm define the search space for the enrichment and are denoted by $\mathcal{X}_{\textrm{sus}}^{(k)}$.
    \item \textbf{Convergence check:} The convergence of the algorithm is then assessed. In this work, we consider convergence to be achieved when the relative variation of the reliability index within two consecutive iterations,
    \begin{equation}\label{eq:conv}
        \varepsilon^{(k)} = \frac{\abs{\beta^{(k)} -  \beta^{(k-1)}}}{\beta^{(k-1)}},
    \end{equation}
    is below a given threshold $\bar{\varepsilon}$ \emph{three times in a row}, where $\beta^{(k)} = - \Phi^{-1}\prt{P_f^{(k)}}$. Here $\Phi^{-1}$ is the inverse standard Gaussian cumulative distribution function and  $\varepsilon$ is a threshold set to $0.005$. If this condition is not respected, the algorithm proceeds with the next step, otherwise Step 10 is performed. Let us note here that common random numbers \citep{Spall2003_ch14} is enabled within different iterations, making the difference within two consecutive iterations mainly due to the actual change in limit-state surface approximation. It is also worth mentioning that this stopping criterion does not guarantee convergence to the underlying true failure probability. However, it has proven to be robust enough for practical purpose and its settings are defined following a large benchmark in \citet{MoustaphaSS2022}.
    \item \textbf{Learning function:} The learning function $U_{\textrm{sys}}$ is then evaluated to find out which points are most likely to improve the accuracy of the system limit state surface, and hence of the failure probability estimate, once added to the experimental design. To this effect, we propose a new system-adapted learning function that is described in Section~\ref{sec:enrichment}.
    \item \textbf{Candidates filtering:} To select the next samples to evaluate, the search space samples $\mathcal{X}_{\textrm{sus}}^{(k)}$ are first filtered into a subset of candidates for enrichment denoted by $\mathcal{X}_{\textrm{sus}}^{\prime (k)}$. This filter simply consists in keeping the most promising samples which correspond to points with the smallest value of $U_{\textrm{sys}}$ as described in Section~\ref{sec:enrichment}. This step allows to avoid possibly costly operations on the whole search space $\mathcal{X}_{\textrm{sus}}^{(k)}$.
    \item \textbf{Samples selection:} The filtered candidates are then clustered using the clustering algorithm \emph{DBSCAN} (density-based spatial clustering with applications to noise), which has the advantage of not requiring a pre-assigned number of clusters. The latter is instead directly inferred from the data. In each identified cluster, the sample with the smallest $U_{\textrm{sys}}$ value is selected for enrichment. The set of selected enrichment samples is denoted by $\mathcal{X}_{\textrm{enr}}^{(k)}$.
    \item \textbf{Sensitivity analysis:} Sensitivity analysis of the composition function $h\prt{z_1, \ldots, z_m}$ is carried out to select the limit states $j \in \acc{1, \ldots, m}$ to evaluate for each enrichment sample. This crucial step, which allows us to only update the  limit states that are the most relevant to system failure, is further detailed in Section~\ref{sec:enrichment}.
    \item \textbf{Experimental design and surrogate updates:} The experimental designs and surrogates which correspond to the limit states that have been selected in the previous step are eventually updated. The algorithm then returns to Step 2.
    \item \textbf{Final estimate:} Once the convergence of the algorithm is reached, a final estimate of the probability of failure is performed using another set-up of subset simulation that allows us to increase the precision of the estimate. It is worth stressing that an overkill set-up is also used within the enrichment iterations ($5 \cdot 10^4$ samples per subset level, intermediate probabilities set to $p_0= 0.25$). Ultimately, we expect the space covered in both the on-line and off-line strategies to be the same.
\end{enumerate}

Each of these steps has been designed so as to enhance the efficiency of the algorithm. We now present in details the most important ingredients.

\subsection{Surrogate models}\label{sec:surrogates}
Surrogate models are an important part of the proposed method as they allow us to reduce the computational burden of the algorithm by providing inexpensive proxies of the limit state functions. The latter are approximated separately using independent experimental designs. While the proposed framework does not rely on a specific type, two different surrogate models are considered in this work, namely Kriging and polynomial chaos-Kriging (PC-Kriging). Both are chosen because they feature built-in prediction errors that can be used in the enrichment scheme. 

\subsubsection{Kriging}
Kriging, also known as Gaussian process modeling, is a surrogate type that considers the underlying function to approximate as one realization of a Gaussian process. This is expressed in the form \citep{Santner2003,Rasmussen2006}
\begin{equation}
	\label{eq:GP}
	g\prt{\ve{x}} = \mu\prt{\ve{x}} + \sigma^2 Z\prt{\ve{x},\omega},
\end{equation}
where $\mu\prt{\ve{x}}$ is the trend, $\sigma^2$ is the Kriging variance and $Z\prt{\ve{x},\omega}$ is a zero-mean unit-variance stationary Gaussian process. The trend can be expressed in many ways but is generally cast as a polynomial of the form $\ve{\beta}^T\ve{f}(\ve{x})$, where $\ve{f} = \acc{f_1, \ldots, f_p}$ is a set of $p$ basis functions and $\ve{\beta} = \acc{\beta_1, \ldots, \beta_p}$ are the corresponding coefficients. The stochastic process $Z\prt{\ve{x},\omega}$, which captures the deviation from the trend, is fully characterized by its auto-correlation function $R\prt{\ve{x},\ve{x}^\prime; \ve{\theta}}$, which is parameterized by $\ve{\theta}$.

In general, we are interested in making a prediction $\widehat{g}$ of the function $g$ at some test location $\ve{x}$ using the observations $\mathcal{G}$ in the experimental design. By definition of a Gaussian random field, any subset of variables including the observations and the response at a given test location are jointly Gaussian, \ie
\begin{equation}
	\begin{Bmatrix}
		\widehat{g}(\ve{x}) \\
		\mathcal{G} 
	\end{Bmatrix}
	\sim
	\mathcal{N}_{N+1}
	\begin{pmatrix}
		\begin{Bmatrix}
			\ve{f}(\ve{x})^T\ve{\beta} \\
			\ve{\ve{F}} \ve{\beta} 
		\end{Bmatrix}
		,
		& 
		\sigma^2
		\begin{Bmatrix}
			1 & \ve{r}^T(\ve{x})  \\
			\ve{r}(\ve{x}) & \ve{R}
		\end{Bmatrix}
	\end{pmatrix},
\end{equation}
where $\ve{F}$ and $\ve{R}$ are the $N \times N$ observation and auto-correlation matrices with elements respectively defined by $F_{ij} = f_j\prt{\ve{x}^{(i)}}$ and $R_{ij} = R\prt{\ve{x}^{(i)},\ve{x}^{(j)};\ve{\theta}}$, $r\prt{\ve{x}}$ is the vector of cross-correlations whose elements are $r_i = R\prt{\ve{x},\ve{x}^{(i)}; \ve{\theta}}, \, i = 1, \ldots N$ and $N$ is the number of samples in the experimental design. 

The prediction is obtained by conditioning the response at the test location with respect to the available observations. This also results in a Gaussian distribution 
\begin{equation}
    \widehat{g}(\ve{x})|\mathcal{G} \sim \mathcal{N}\prt{\widehat{\mu}_{\widehat{g}}(\ve{x}), \widehat{\sigma}_{\widehat{g}}^2(\ve{x})},
\end{equation}
whose mean and variance are respectively given by
\begin{equation}\label{eq:KRGmean}
\mu_{\widehat{g}}\prt{\ve{x}}=\ve{f}^T\prt{\ve{x}} \widehat{\ve{\beta}}+\ve{r}^T\prt{\ve{x}}\ve{R}^{-1} \big(\mathcal{G}-\ve{F}\widehat{\ve{\beta}} \big),
\end{equation}
\begin{equation}\label{eq:KRGvar}
\sigma^2_{\widehat{g}}\prt{\ve{x}}=\widehat{\sigma}^2 \prt{1- \ve{r}^T\prt{\ve{x}} \ve{R}^{-1} \ve{r}\prt{\ve{x}} + \ve{u}^T\prt{\ve{x}}(\ve{F}^T\ve{R}^{-1}\ve{F})^{-1} \ve{u}\prt{\ve{x}} }.
\end{equation}
In these two equations, $\widehat{\ve{\beta}} = \prt{\ve{F}^T\ve{R}^{-1}\ve{F}}^{-1} \ve{F}^T \ve{R}^{-1} \mathcal{G}$ is the least-square estimate of the regression coefficients, $\widehat{\sigma}^2 = \frac{1}{N} \prt{\mathcal{G}- \ve{F}\ve{\beta}}^T \ve{R}^{-1} \prt{\mathcal{G}- \ve{F}\ve{\beta}}$ is the maximum likelihood estimate of the process variance and $\ve{u}\prt{\ve{x}} = \ve{F}^T \ve{R}^{-1} \ve{r}\prt{\ve{x}} - \ve{f}\prt{\ve{x}}$ has been introduced for convenience.

An important feature of Kriging is that besides prediction, a measure of its own accuracy is also readily provided through the variance in Eq.~\eqref{eq:KRGvar}. It generally tends to get large further away from the training points and collapses to zero on them, thus making Kriging an interpolant. This feature is essential in active learning strategies.

\subsubsection{Polynomial Chaos-Kriging}
Polynomial Chaos-Kriging or PC-Kriging attempts to combine the global behavior of polynomial chaos expansions (PCE, \citet{Xiu2002}) to the local behavior of Kriging. It is built as a universal Kriging model where the trend is actually a PCE model \citep{SchoebiIJUQ2015,SchoebiASCE2017}:
\begin{equation}
	\label{eq:PCK}
	g\prt{\ve{x}} = \sum_{\ve{\alpha} \subset \mathcal{A}} y_{\ve{\alpha}} \Psi_{\ve{\alpha}} \prt{\ve{x}} + \sigma^2 Z\prt{\ve{x},\omega},
\end{equation}
where $\Psi_{\ve{\alpha}}\prt{\ve{x}}$ are multivariate polynomials orthonormal with respect to the input distribution $f_{\ve{X}}$, $\ve{\alpha} \in \mathcal{A} \subset \mathbb{N}^M$ are multi-indices identifying the components of the multivariate components and $y_{\ve{\alpha}}$ are corresponding regression coefficients.

The multivariate polynomials are assembled using a tensor product of their univariate  counterparts, \emph{i.e.},
\begin{equation}
    \Psi_{\ve{\alpha}}\prt{\ve{x}} = \prod_{i=1}^M \Psi_{\alpha_i}\prt{x_i}.
\end{equation}
 The optimal basis is obtained here using least-angle regression (LAR), which allows one to adaptively select basis functions that lead to a sparse representation \citep{Efron2004,Blatman2011}. 
 Once the basis is selected, the hyperparameters   $\acc{y_{\ve{\alpha}},\,\sigma^2, \ve{\theta}}$ are estimated using the traditional Kriging calibration method described in the previous section (See \citet{SchoebiIJUQ2015,UQdoc_20_109} for more details).

\subsection{Enrichment scheme}\label{sec:enrichment}
To compute the probability of failure at each iteration of the active learning process, we consider subset simulation \citep{Au2001}. Compared to crude Monte Carlo, subset simulation allows us to more thoroughly explore the random variable space and generally provides us with finer reliability estimates at a lower cost. 

In this proposed framework, reliability estimation is performed at each iteration using the approximated system limit state $h\prt{\widehat{g}_1\prt{\ve{x}}, \ldots, \widehat{g}_m\prt{\ve{x}}}$. The samples generated by subset simulation, herein denoted by $\mathcal{X}_{\textrm{sus}}$, are the starting point of the enrichment process as they constitute the candidate pool from which samples to add to the experimental design are selected. Three components are leveraged to optimize the necessary number of samples to convergence. They are now described in details. 

\paragraph{Learning function\\}
The learning function is an important component of active learning as it allows us to select the next point(s) to add to the experimental design. It generally consists of a simple function that is maximized (or minimized) to yield the point that is the most likely to improve the knowledge of the limit state surface, and henceforth the accuracy of the surrogate-based estimate of the failure probability. 

There are many such learning functions in the literature and in this work we propose an adaptation of the well-known deviation number \citep{Echard2011}. Considering a Gaussian process approximation of the limit state function $\widehat{g}\prt{\ve{x}} \sim \mathcal{N}\prt{\mu_{\widehat{g}}\prt{\ve{x}}, \, \sigma_{\widehat{g}}^2\prt{\ve{x}}}$, the deviation number simply reads:
\begin{equation}
    U\prt{\ve{x}} = \frac{\abs{\mu_{\widehat{g}}\prt{\ve{x}}}}{\sigma_{\widehat{g}}\prt{\ve{x}}}.
\end{equation}
The next point to add to the experimental design is then the one that minimizes $U$. Minimal values of the learning function are obtained when the GP prediction is close to the $0$ (points in the vicinity of the limit state surface) or when the prediction variance is large (in other words, points in areas where the prediction sign is most uncertain).

In this work, a \emph{system-adapted deviation number} is considered using the system limit state surrogate given by
\begin{equation}\label{eq:Zsys}
    Z_{\textrm{sys}}\prt{\ve{x}} \stackrel{\text{def}}{=} h\prt{\ve{Z}} = h\prt{\widehat{g}_1\prt{\ve{x}_1}, \ldots, \widehat{g}_m\prt{\ve{x}_m}},
\end{equation}
where $Z_j \stackrel{\text{def}}{=} \widehat{g}_j\prt{\ve{x}_j}, j = 1,\ldots,m$ represents the Kriging or PC-Kriging approximation of the $j$-th limit state function. Note that $Z_{\textrm{sys}}$ is a random variable attached to each point $\ve{x}$, obtained by inserting $m$ Gaussian variables $Z_1, \ldots, Z_m$ into the composition function $h$. Note that it is in general \emph{not} Gaussian.

This proposed learning function is then defined by:
\begin{equation}\label{eq:Usys}
    U_{\textrm{sys}}\prt{\ve{x}} = \frac{\abs{\mu_{\textrm{sys}}\prt{\ve{x}}}}{\sigma_{\textrm{sys}}\prt{\ve{x}}},
\end{equation}
where $\mu_{\textrm{sys}}\prt{\ve{x}}$ and $\sigma_{\textrm{sys}}\prt{\ve{x}}$ are respectively the mean and standard deviation of the system response $Z_{\textrm{sys}}\prt{\ve{x}}$. These two quantities are not readily available as the distribution of $Z_{\textrm{sys}}$ is not generally known for a general system and related composition function $h$. It is actually not possible to derive them analytically, even when the (Gaussian) distributions of  $\acc{\widehat{g}_j\prt{\ve{x}_j}, j = 1, \ldots, m}$ are known and the analytical expression of $h$ is given.

We therefore resort to an empirical estimation based on the Kriging/PC-Kriging approximations. In this context, the vector $\ve{Z}$ in Eq.~\eqref{eq:Zsys} is an independent multivariate Gaussian since each of its components is obtained from an independent Kriging/PC-Kriging predictor, \ie, $Z_j \sim \mathcal{N}\prt{\mu_{\widehat{g}_j}\prt{\ve{x}_j}, \, \sigma_{\widehat{g}_j}^2\prt{\ve{x}_j}}$ and
\begin{equation}\label{eq:fz}
    \ve{Z}\prt{\ve{x}} \sim f_{\ve{Z}} = \prod_{j=1}^{m} \mathcal{N}\prt{\mu_{\widehat{g}_j}\prt{\ve{x}_j}, \, \sigma_{\widehat{g}_j}^2\prt{\ve{x}_j}}.
\end{equation}

The empirical estimate of $U_{\textrm{sys}}$ can therefore be obtained by drawing $n$ samples $\acc{\ve{z}^{(1)}, \ldots \ve{z}^{(n)}}$ following Eq.~\eqref{eq:fz} and then estimating the mean and variance of the system response as follows
\begin{equation}\label{eq:EmpMuSigma}
    \begin{split}
        \widehat{\mu}_{\textrm{sys}} = & \frac{1}{n} \sum_{i=1}^{n} h \prt{\ve{z}^{(i)}}, \\
        \widehat{\sigma}_{\textrm{sys}} = & \frac{1}{n-1} \sum_{i=1}^{n} \prt{ h\prt{\ve{z}^{(i)}} - \widehat{\mu}_{\textrm{sys}} }^2, 
    \end{split}
\end{equation}
which are then plugged in Eq.~\eqref{eq:Usys} to get
\begin{equation}\label{eq:Usyshat}
    \widehat{U}_{\textrm{sys}}\prt{\ve{x}} = \frac{\abs{\widehat{\mu}_{\textrm{sys}}}}{\widehat{\sigma}_{\textrm{sys}}}.
\end{equation}
Similarly to the original $U$ function, samples  that minimize $U_{\textrm{sys}}$ are of interest, since they correspond to regions where the system prediction is close to zero and/or where the uncertainty in the prediction is large.

\paragraph{Filtering and clustering\\}
Once the learning function is computed for all the samples in the candidate pool $\mathcal{X}_{\textrm{sus}}$, the latter is filtered so as to keep only the most relevant samples \emph{i.e.}, the ones such that $\widehat{U}_{\textrm{sys}}\prt{\ve{x}}$ (which is non negative by construction) is close to zero. More specifically, the \emph{reduced enrichment candidate set} is defined by
\begin{equation}
    \mathcal{X}^\prime_{\textrm{sus}} = \acc{\ve{x} \in \mathcal{X}_{\textrm{sus}}: U_{\textrm{sys}}\prt{\ve{x}} \leq U_q},
\end{equation}
where $U_q$ is the $\alpha$-th empirical quantile of the set  $U_{\textrm{sys}}\prt{\mathcal{X}_{\textrm{sus}}}$. The quantile level $\alpha$ is chosen in the range $\bra{0.01 - 0.05}$ so as to only keep the most important samples.

Following this filtering step, the samples are clustered into $C$ subsets. This allows us to add multiple points per iteration by possibly considering different failure modes. This also contributes to reducing information redundancy, since points that are eventually added are not too close to each other. Moreover, the number of clusters is not predefined but directly inferred from the data. 
In some cases, each cluster corresponds to a different failure mode of the system. 
There are many ways to detect such clusters and we consider here the so-called \emph{density-based spatial clustering of applications with noise} a.k.a. DBSCAN, which is a hierarchical density-based clustering method proposed by \citet{Ester1996}. The algorithm is described in details in Section~\ref{app:dbscan}. 

Assuming that $C$ clusters $\acc{\mathcal{X}_c, c = 1 \enum C}$ have been identified, one enrichment point is defined for each of them as follows:
\begin{equation}\label{eq:minUsys}
    \widetilde{\ve{x}}^{(c)} = \arg \min_{\ve{x} \in \mathcal{X}_c} U_{\textrm{sys}}\prt{\ve{x}}, \, c = 1, \ldots, C.
\end{equation}
The resulting set of enrichment points is denoted in the sequel by $\mathcal{X}_{\textrm{enr}}$. Enriching all the experimental designs with these points is not optimal as each of them is rather connected to a specific failure mode. We therefore consider \emph{sensitivity analysis} of the composition function $h$ to identify for each sample which limit state is the most impacted. 

\paragraph{Sensitivity analysis\\}
Sensitivity analysis is used to identify, for each enrichment sample, which limit state contributes the most to system failure. In this work, we consider the popular Sobol' sensitivity indices \citep{Sobol1993,Iooss2022}.

Let us consider one sample $\widetilde{\ve{x}}^{(c)} \in \mathcal{X}_{\textrm{enr}}$. 
We can calculate the Sobol' sensitivity indices of the function $h\prt{\widehat{g}\prt{\widetilde{\ve{x}}^{(c)}}} = h\prt{\ve{Z}\prt{\widetilde{\ve{x}}^{(c)}}}$ with respect to the random variables $\acc{Z_1\prt{\widetilde{\ve{x}}^{(c)}}, \ldots, Z_m\prt{\widetilde{\ve{x}}^{(c)}}}$. 
These indices can be computed by Monte Carlo simulation since $h$ is an inexpensive function and samples from the Gaussian random vector $\ve{Z}\prt{\widetilde{\ve{x}}^{(c)}}$ can be easily obtained, given that they are independent multivariate Gaussian whose parameters are provided by the Kriging/PC-Kriging models, \ie,
\begin{equation}
Z_j\prt{\widetilde{\ve{x}}^{(c)}} = \widehat{g}_j\prt{\widetilde{\ve{x}}_{j}^{(c)}}\sim \mathcal{N}\prt{\mu_{\widehat{g}_j}\prt{\widetilde{\ve{x}}_j^{(c)}}, \, \sigma_{\widehat{g}_j}^2\prt{\widetilde{\ve{x}}_j^{(c)}}}, \, j = 1, \ldots, m
\end{equation}

The total Sobol' indices $\acc{S_1^T\prt{\widetilde{\ve{x}}^{(c)}},\ldots, S_m^T\prt{\widetilde{\ve{x}}^{(c)}}}$ are chosen to rank the limit state in terms of relevance for each enrichment sample $\widetilde{\ve{x}}^{(c)}$. The limit state to enrich is therefore selected as the one with the corresponding largest total Sobol' index, \emph{i.e.,}
\begin{equation}
    j^\ast\prt{\widetilde{\ve{x}}^{(c)}} = \arg \max_{j = 1 \ldots m} S_j^T\prt{\widetilde{\ve{x}}^{(c)}}.
\end{equation}
In summary, for the $c$-th enrichment point, the experimental design $\mathcal{D}_{j^\ast}$ is updated such that $\mathcal{D}_{j^\ast} \leftarrow \mathcal{D}_{j^\ast} \cup \prt{\widetilde{\ve{x}}_{j^\ast}^{(c)}, g\prt{\widetilde{\ve{x}}_{j^\ast}^{(c)}}}$, where $\widetilde{\ve{x}}_{j^\ast}^{(c)}$ is the relevant subset of $\widetilde{\ve{x}}^{(c)}$ corresponding to the $j^\ast$-th limit state.

\paragraph{Summary of the enrichment scheme\\}
The entire enrichment process at a given iteration is summarized in Algorithm~\ref{Alg:ENR}. The superscript $k$, which refers to the iteration number of the active learning process is omitted to simplify the notation.

\begin{algorithm}[!ht]
	\caption{Enrichment process}
	\begin{algorithmic}[1]
		\Require{}
		\Statex Get the samples $\mathcal{X}_{\textrm{sus}}$ generated by subset simulation
		\Statex Get the current limit state surrogates $\widehat{g}_1, \ldots, \widehat{g}_m$
		\Statex Get the current experimental designs $\mathcal{D}_1, \ldots, \mathcal{D}_m$
		\Statex Set the quantile level $\alpha$ for filtering \Comment{\color{DarkBlue} {\scriptsize{\emph{e.g.}, $\alpha = 0.01$ }} \color{black}}
		\Statex Set the maximum number of enrichment points $N_{\textrm{max}}$ \Comment{\color{DarkBlue} {\scriptsize{\emph{e.g.}, $N_{\textrm{max}} = m$ }} \color{black}}
		\Statex \hrulefill%
		\For{each $\ve{x} \in \mathcal{X}_{\textrm{sus}} $}
		\State Draw $n$ samples $\acc{\ve{z}^{(1)},\ldots, \ve{z}^{(n)}}$ following $f_{\ve{Z}}$ \Comment{\color{DarkGreen} {\scriptsize{$f_{\ve{Z}} = \prod_{j=1}^{m}\mathcal{N}\prt{\mu_{\widehat{g}_j}\prt{\ve{x}_j}, \, \sigma_{\widehat{g}_j}^2\prt{\ve{x}_j}}$ }}  \color{black}}
		\State Calculate  $\widehat{\mu}_{\textrm{sys}}$ and $\widehat{\sigma}_{\textrm{sys}}$ following  Eq.~\eqref{eq:EmpMuSigma}
		\State Calculate $\widehat{U}_{\textrm{sys}}\prt{\ve{x}}$ following Eq.~\eqref{eq:Usyshat}
		\EndFor
		\State Calculate $U_q$, the $\alpha$-quantile of $\widehat{U}_{\textrm{sys}}\prt{\mathcal{X}_{\textrm{sus}}}$
		\State Define the reduced candidate set $\mathcal{X}^\prime_{\textrm{sus}}$  \Comment{\color{DarkGreen} {\scriptsize{$\mathcal{X}^\prime_{\textrm{sus}} = \acc{\ve{x} \in \mathcal{X}_{\textrm{sus}}: U_{\textrm{sys}}\prt{\ve{x}} \leq U_q}$ }}  \color{black}}
		\State Partition $\mathcal{X}^\prime_{\textrm{sus}}$ into $C$ clusters $\acc{\mathcal{X}_1, \ldots, \mathcal{X}_C}$\Comment{\color{DarkGreen} {\scriptsize{$C$ is automatically detected by DBSCAN}} \color{black}}
  \If{$C > N_{\textrm{max}}$} 
  \State Only keep the $N_{\textrm{max}}$ samples with smallest $U_{\textrm{sys}}$ values
  \EndIf
		\State Get the enrichment samples $\mathcal{X}_{\textrm{enr}} = \acc{\widetilde{\ve{x}}^{(1)} \enum  \widetilde{\ve{x}}^{(C)}}$ using Eq.~\eqref{eq:minUsys}
		\For{each $\widetilde{\ve{x}}^{(c)} \in \mathcal{X}_{\textrm{enr}}$}
		\State Calculate the total Sobol' indices $\acc{S_1^T, \ldots, S_m^T}$ of $h\prt{u_1, \ldots, u_m}$ where $u_j =\ve{Z}_j\prt{\widetilde{\ve{x}}^{(c)}}$
		\State Find the component $j^\ast$ with the largest index $S_{j^\ast}^T$
		\State Update the corresponding experimental design $\mathcal{D}_{j^\ast} \leftarrow \mathcal{D}_{j^\ast} \cup \prt{\widetilde{\ve{x}}_{j^\ast}^{(c)}, g\prt{\widetilde{\ve{x}}_{j^\ast}^{(c)}}}$
  \State Update the corresponding surrogate model $\widehat{g}_{j^\ast}$
		\EndFor
	\end{algorithmic}
	\label{Alg:ENR}
\end{algorithm}


\section{Application examples}\label{sec:Examples}
We now consider three different problems to illustrate and validate the proposed approach. The first two are based on analytical functions and will be used to benchmark our method against the recent literature. The third example is a finite element model of a system of electrical towers used to showcase the applicability and efficiency of our method on a complex high-dimensional problem.

To assess the accuracy of the proposed method, we consider the following relative error for each example:
\begin{equation}
    \varepsilon_{\beta} = \frac{\abs{\beta - \bar{\beta}}}{\bar{\beta}},
\end{equation}
where $\beta = -\Phi^{-1}\prt{P_f}$ is the reliability index obtained by a surrogate-assisted approach  and $\bar{\beta}$ is the reference solution. The latter is computed using subset simulation tuned such that the resulting coefficient of variation is lower than $0.005$.

In active learning, the goal is to find the probability of failure using a limited number of calls to the original model. Therefore, we consider the number of model evaluations $N_{\textrm{eval}}$ as criterion of efficiency. Finally, we assess the robustness of the proposed approach by repeating each analysis  $15$ times using different initial experimental designs.

Both Kriging and PC-Kriging are considered in the two benchmark examples. 
For Kriging, we consider a linear trend while for PC-Kriging the maximum polynomial degree  of the PCE part is set equal to $3$. The convergence threshold in Eq.~\eqref{eq:conv} is set to $\bar{\varepsilon} = 5 \cdot 10^{-3}$.  
The size of the initial experimental design is defined as $N_j^{(0)} = 2 M_j + 1$, where $M_j$ is the effective input dimensionality of the $j$-th component limit state. 
Regarding the enrichment scheme, the quantile level for filtering $\mathcal{X}_{\textrm{sus}}$ into $\mathcal{X}^\prime_{\textrm{sus}}$ is set equal to $\alpha = 0.01$, while the maximum number of enrichment points allowed per iteration is set to $N_{\textrm{max}} = M$.

\subsection{Four-branch function}
This example is popular in the active learning community as it has been used often for benchmarks \citep{Echard2011}. 
It is a two-dimensional problem, which makes it ideal to illustrate different aspects of the proposed method graphically. We consider a series system which consists of $m=4$ component limit states defined as \citep{Waarts2000}:
\begin{equation}
\begin{cases}
    g_1(\ve{X}) = 3 + 0.1(X_1-X_2)^2 - \frac{1}{\sqrt{2}}(X_1+X_2), \\
    g_2(\ve{X}) = 3 + 0.1(X_1-X_2)^2 + \frac{1}{\sqrt{2}}(X_1+X_2), \\
    g_3(\ve{X}) = (X_1-X_2) + \frac{P}{\sqrt{2}}, \\
    g_4(\ve{X}) = (X_2-X_1) + \frac{P}{\sqrt{2}},
    \end{cases}
\end{equation}
where $X_1, X_2 \sim \mathcal{N}\prt{0,1}$ and $P$ is a parameter that has been set to either $6$ or $7$ in the literature. We consider both values for benchmarking purposes but all illustrations correspond to the case $P=7$. The reference probability of failure for $P=7$ (resp. $P=6$) is $2.24\cdot 10^{-3}$ (resp. $4.48\cdot 10^{-3}$). This corresponds to a reliability index of $2.84$ (resp. $2.61$).

\begin{figure}[!ht]
	\centering
	\subfloat[Initial ED and limit state]{\label{fig:Ex1:Init_a}\includegraphics[width=0.49\textwidth]{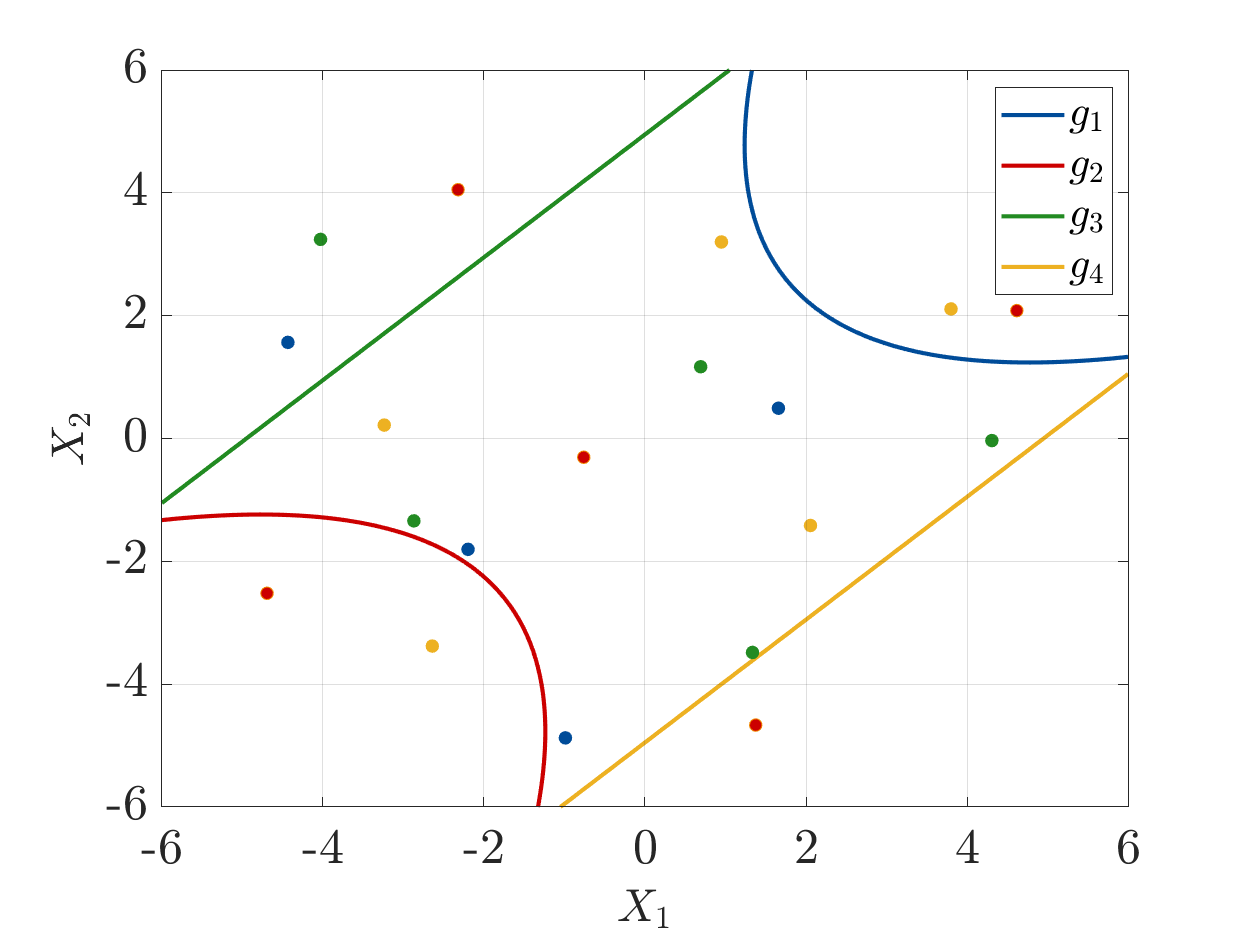}}%
	\subfloat[Candidate pool $\mathcal{X}_{\textrm{sus}}$]{\label{fig:Ex1:Init_b}\includegraphics[width=0.49\textwidth]{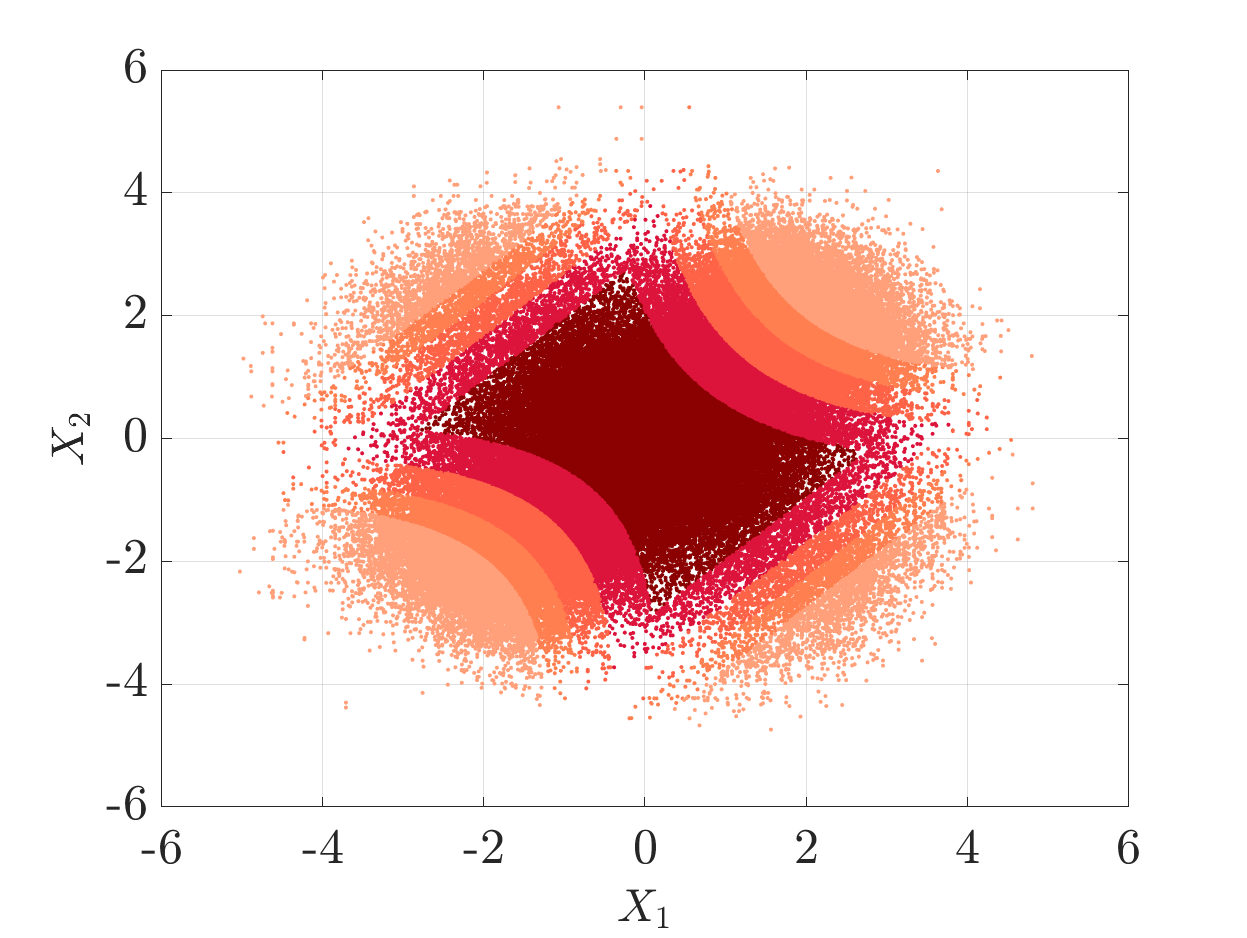}}%
	\caption{Example 1 - limit state functions, initial experimental designs and candidate
    pool for enrichment of the four-branch function. In the left panel, the colors of the samples correspond to the limit-state they are associated with. The color shades in the right panel show the different steps of subset simulation.}
	\label{fig:Ex1:Init}
\end{figure}

Let us start by illustrating the enrichment process. Figure~\ref{fig:Ex1:Init_a} shows one realization of the initial experimental design together with the limit state surfaces (5 points for each limit state). Different colours correspond to different components. The limit state functions $g_1$ and $g_2$ are nonlinear while $g_3$ and $g_4$ are linear, and hence easier to approximate. Regarding the initial ED, the use of the hypercube allows us to spread the samples in the input space and reach the failure domain at initialization. This further reduces the chance of premature convergence where one of the failure domains is not discovered.

\begin{figure}[!ht]
	\centering
	\subfloat[Iteration $\#1$]{\label{fig:Ex1:Xenr_a}\includegraphics[width=0.45\textwidth]{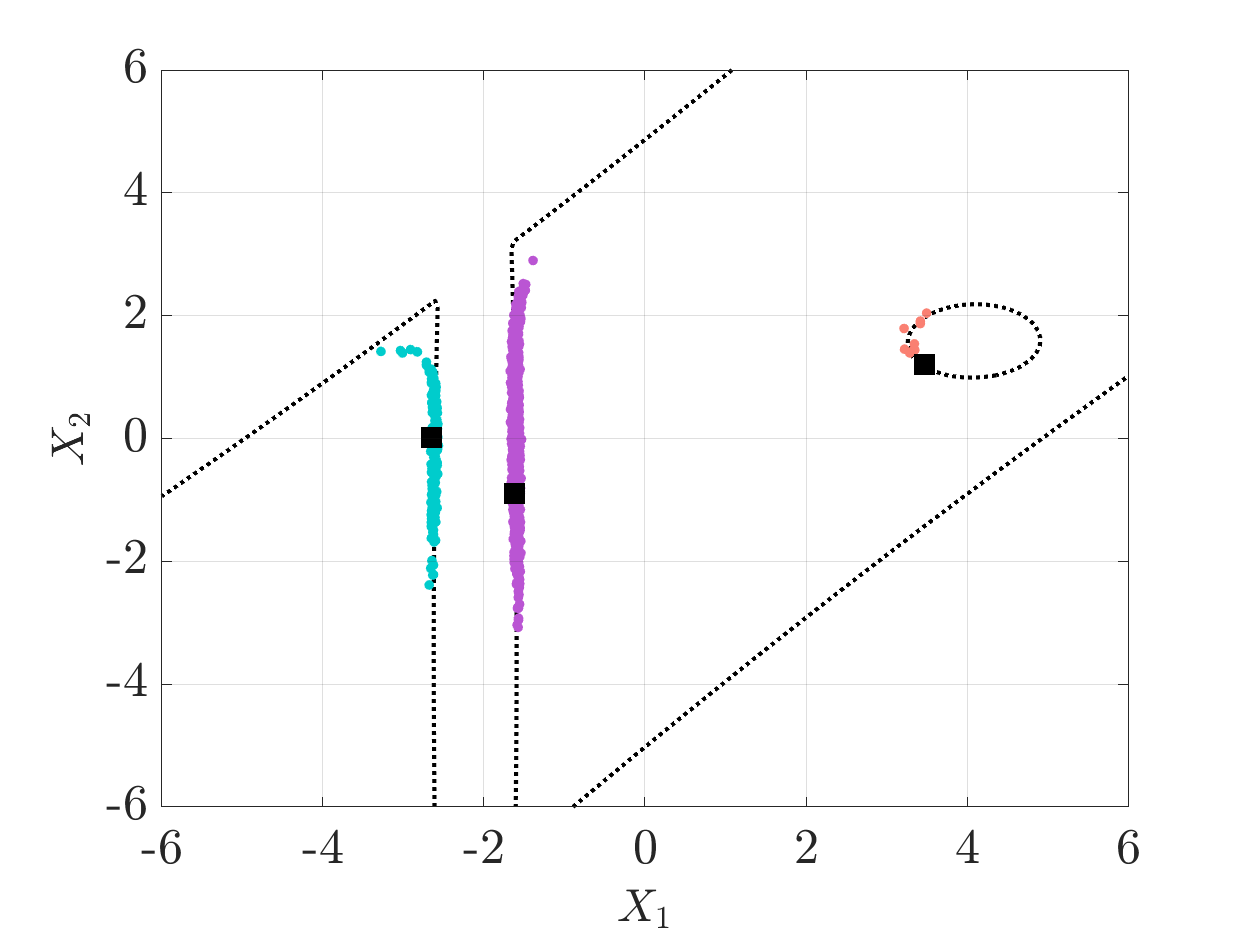}}%
	\subfloat[Iteration $\#4$]{\label{fig:Ex1:Xenr_b}\includegraphics[width=0.45\textwidth]{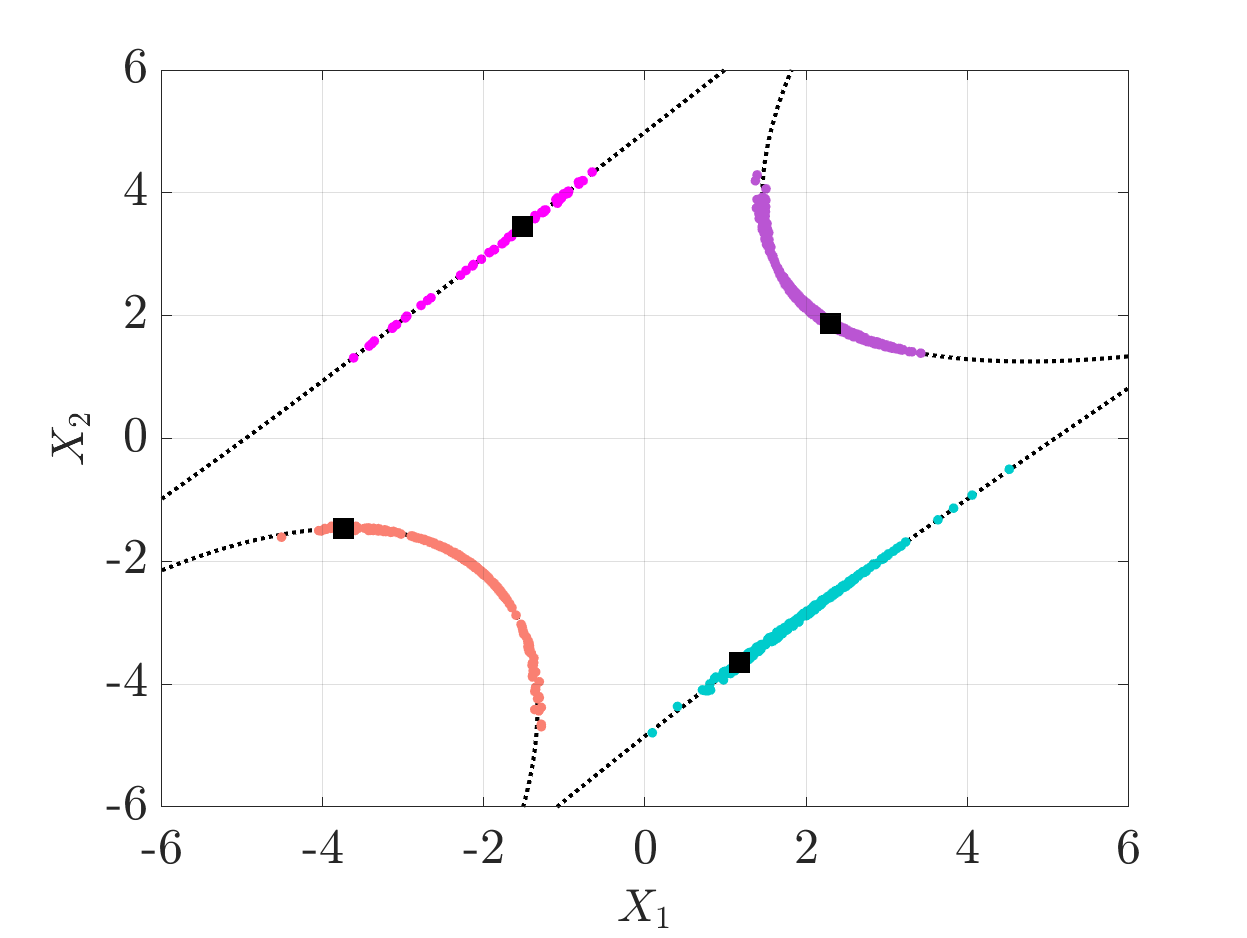}}%
	\caption{Example 1 - limit state functions and enrichment samples (black dots) using ordinary Kriging. The colored dots show the clustered  enrichment candidates.}
	\label{fig:Ex1:Xenr}
\end{figure}

Figure~\ref{fig:Ex1:Init_b} shows the candidate pool for one enrichment iteration $\mathcal{X}_{\textrm{sus}}$, which are the samples generated by subset simulation. 
The different shades of colours represent the different subset iterations. These subsets ensure a higher likelihood of reaching the failure domain compared to crude Monte Carlo simulation. Those samples are then filtered so as to keep a small fraction with the smallest $U_{\textrm{sys}}$. This is illustrated in Figure~\ref{fig:Ex1:Xenr} for two different iterations when using ordinary Kriging. In these figures, the dotted lines represent the system limit state surface approximation and the coloured samples represent the filtered enrichment samples $\mathcal{X}_{\textrm{sus}}^\prime$. Each colour corresponds to a different cluster. As it can be seen, the number of identified clusters depends on the current state of the limit state surfaces. At iteration $\#1$ (Figure~\ref{fig:Ex1:Xenr_a}), three (erroneous) failure modes are identified due to the small ED size that lead to inaccurate surrogate models. Consequently, the enrichment set $\mathcal{X}_{\textrm{enr}}$ contains three samples which are represented by the black squares. In contrast at iteration $\#4$ (Figure~\ref{fig:Ex1:Xenr_b}), the surrogate models are more accurate and the four failure modes are properly identified.

\begin{figure}[!ht]
	\centering
	\subfloat[Kriging]{\label{fig:Ex1:enr_a}\includegraphics[width=0.45\textwidth]{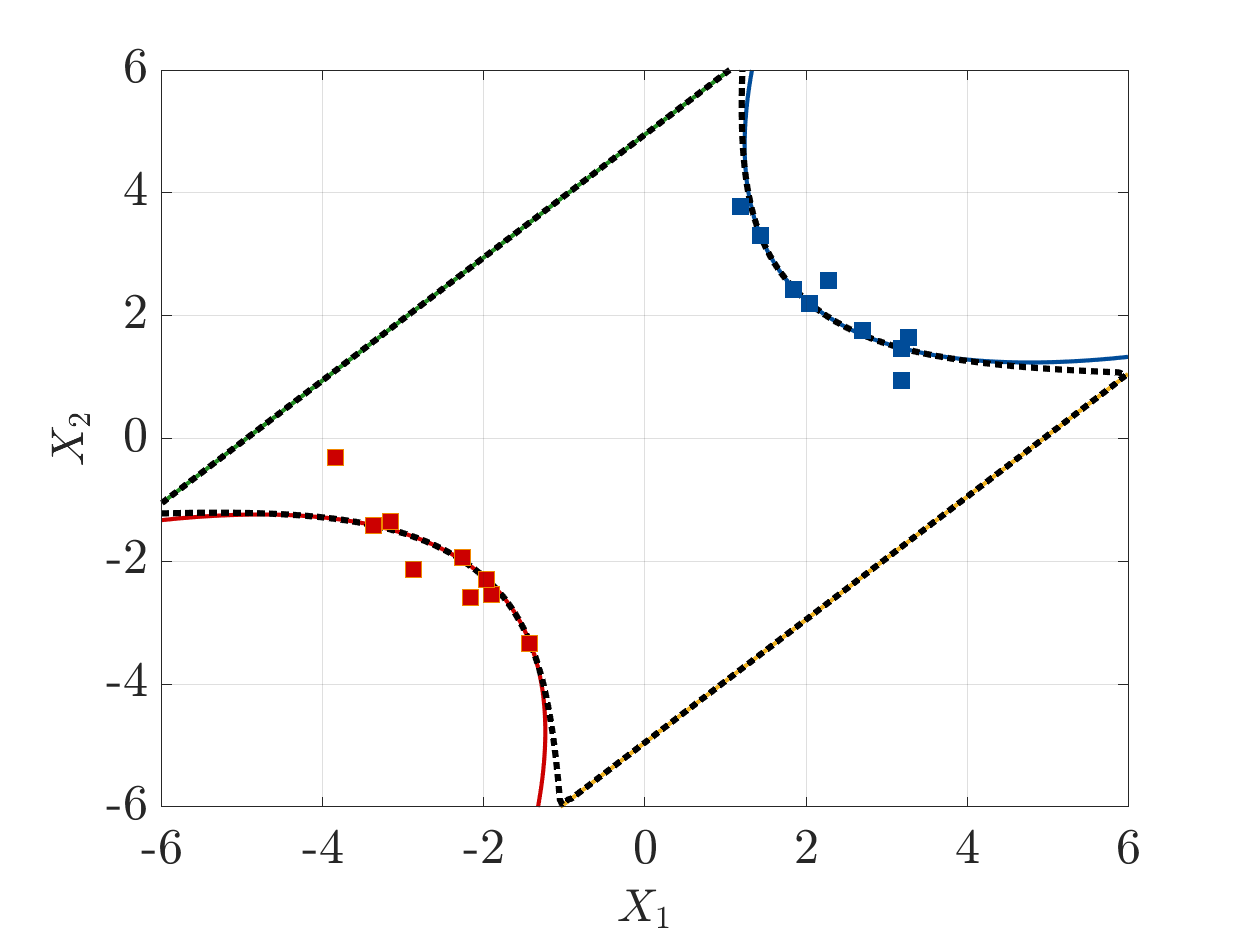}}%
	\subfloat[PC-Kriging]{\label{fig:Ex1:enr_b}\includegraphics[width=0.45\textwidth]{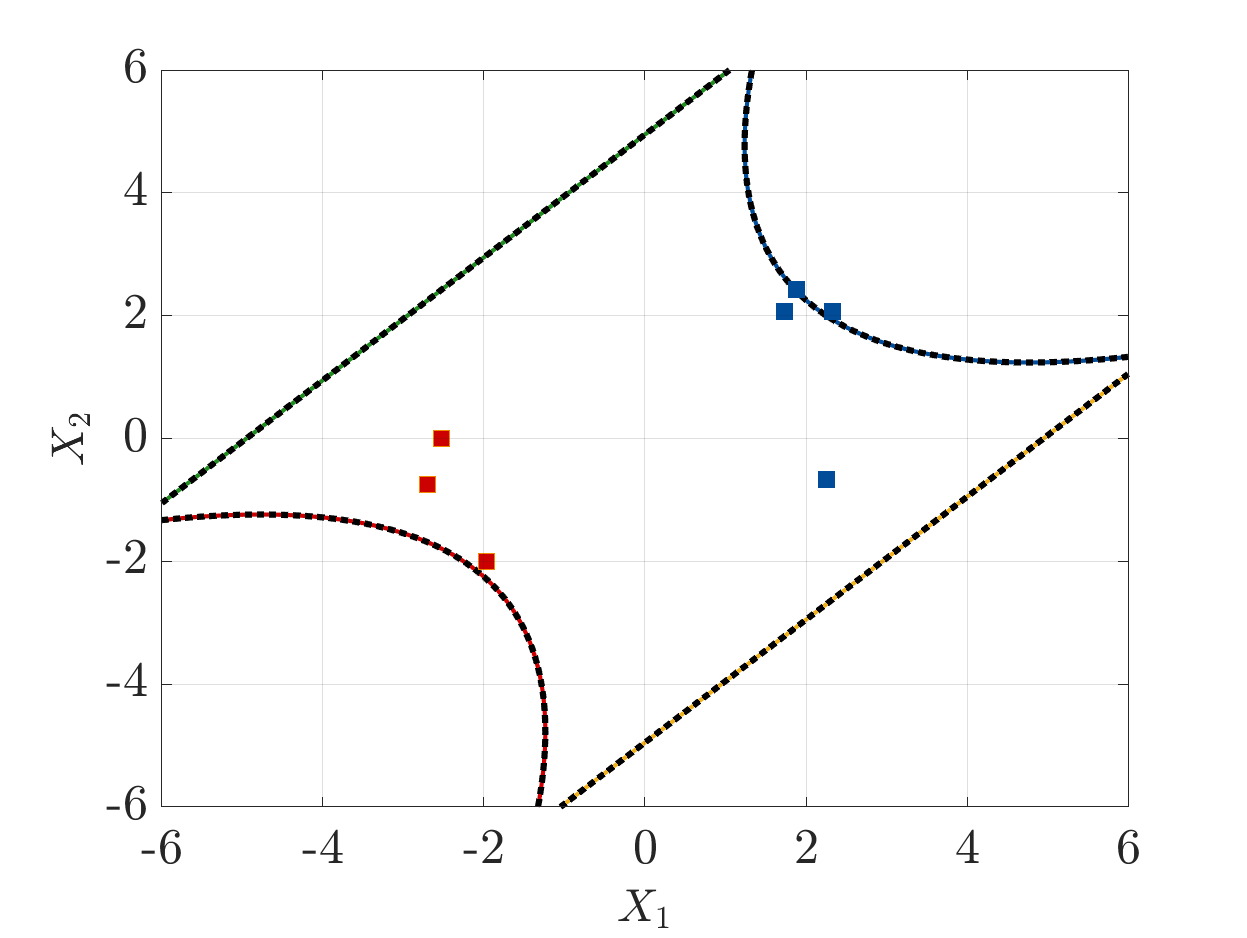}}%
	\caption{Example 1 - Enrichment samples for the median case with $P=7$.}
	\label{fig:Ex1:enr}
\end{figure}

The final set of enrichment samples are illustrated for the median case in Figure~\ref{fig:Ex1:enr} for both Kriging and PC-Kriging. The different colours show which limit state was actually evaluated for each sample. It is worth noting here that the linear limit-states $g_3$ and $g_4$ are never enriched when using universal Kriging with linear trend or PC-Kriging as they are very well approximated, even when the ED is extremely small. 

\begin{figure}[!ht]
	\centering
	\subfloat[Reliability index]{\label{fig:Ex1:Results_a}\includegraphics[width=0.45\textwidth]{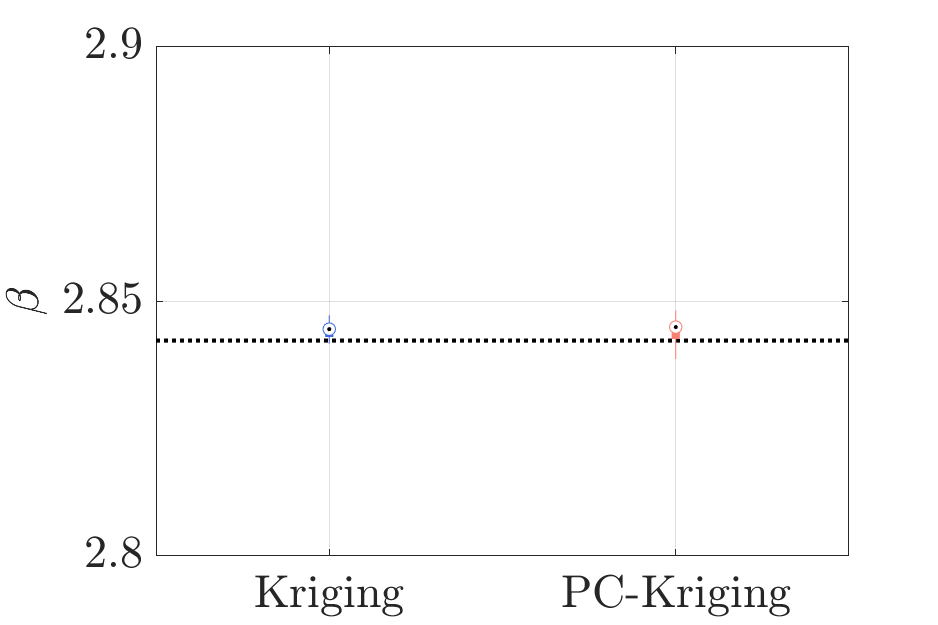}}%
	\subfloat[Number of model evaluations]{\label{fig:Ex1:Results_b}\includegraphics[width=0.45\textwidth]{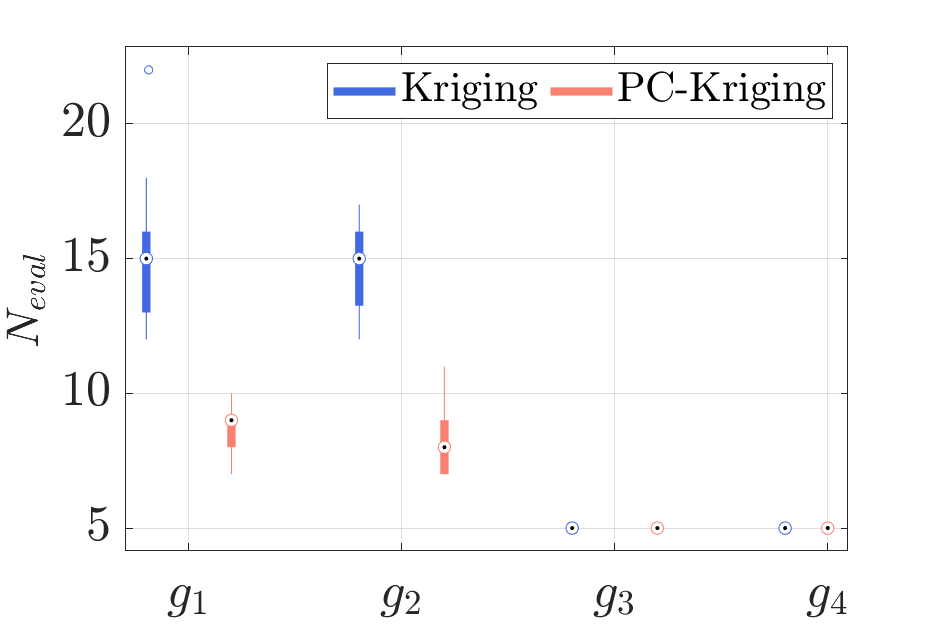}}%
	\caption{Example 1 - Boxplots summarizing the results of the $15$ repetitions for $P=7$.}
	\label{fig:Ex1:Results}
\end{figure}

Figure~\ref{fig:Ex1:Results} summarizes the  results of the analysis considering all $15$ repetitions. In details, Figure~\ref{fig:Ex1:Results_a} shows boxplots of the reliability index when using Kriging or PC-Kriging.  In the boxplot, the circled dot represents the median solution, the thick line corresponds to the first and third quartiles while the thin whiskers show the range of all solutions, outliers excluded. The horizontal dotted black line corresponds to the reference solution. It can be seen that the proposed framework is extremely accurate in both cases and robust, as very little dispersion can be observed. Figure~\ref{fig:Ex1:Results_b} shows boxplots of the number of models evaluations for each component. As expected limit states $g_1$ and $g_2$ which are the more difficult to approximate and which contribute the most to system failure are evaluated more often than $g_3$ and $g_4$. As a matter of fact, the latter are very well approximated by both Kriging and PC-Kriging thanks to the polynomial trend and are therefore never enriched in these replications.

We now compare our results to other methods available in the literature. The case $P=7$ was mainly used with component reliability where the system limit state $h\prt{\ve{g}\prt{\ve{x}}}$ was directly approximated. Table~\ref{tab:Ex1:comp} shows results gathered by \citet{Ling2019} together with the median results obtained from the repetitions using our proposed method (with Kriging and PC-Kriging). As shown by the error column $\varepsilon_{\beta}$, all methods are accurate.  However, our method is much more efficient as can be seen from the required number of model evaluations to convergence. The methods from the literature considered all the four limit state function as one, hence the factor $4$ in the $N_{\textrm{eval}}$ column. 
Yet, even without multiplying by four, the methods we propose remain much more efficient. 
When using Kriging, we converge in $40$ model evaluations (median value), which is much smaller than the $61$ ($\times 4$) obtained by \citet{Ling2019}. 
When using PC-Kriging, the convergence is even faster. 
In total, only $27$ samples are required to yield an accurate estimate of the probability of failure.
\begin{table}[!ht]
\caption{Four-branch function ($P=7$) performance comparison from \citet{Ling2019}.}
\label{tab:Ex1:comp}
    \centering
    \begin{tabular}{l c c c c}
    \hline 
         Method & $P_f$ & $\beta$ & $\varepsilon_{\beta}$ &  Total $N_{\text{eval}}$\\
         \hline 
         SuS (reference) & $2.239 \cdot 10^{-3}$ & $2.842$ &  $--$ & $\sim 4 \cdot 10^5$ \\
         AK-MCS & $2.271 \cdot 10^{-3}$ & $2.838$  &  $1.524 \cdot 10^{-3}$ &  $102 \times 4$ \\
         AK-IS & $2.210 \cdot 10^{-3}$ &  $2.847$ & $1.643 \cdot 10^{-3}$ &  $361.5 \times 4$ \\
         AK-SS & $2.2480 \cdot 10^{-3}$ & $2.841$ &  $4.684 \cdot 10^{-4}$ &  $67 \times 4$ \\
         Ling & $2.2480 \cdot 10^{-3}$ & $2.841$ &  $4.684 \cdot 10^{-4}$ &  $61.4 \times 4$ \\
         Proposed (KRG) & $2.224 \cdot 10^{-3}$ & $2.845$ & $7.861 \cdot 10^{-4}$ & $40$ \\
         Proposed (PCK) & $2.221 \cdot 10^{-3}$ & $2.845$ & $9.285 \cdot 10^{-4}$ & $27$ \\
         \hline
    \end{tabular}
\end{table}

To compare our method with dedicated system reliability methods, we consider the case when $P=6$. Table~\ref{tab:Ex1:sys} shows results gathered by \citet{Guo2021} together with our proposed method. The benchmark  methods are this time much more efficient than the ones used above for $P=7$, since they were precisely developed to solve system problems. They require on average $62$ model evaluations. This makes them still largely less efficient than the method we propose, which converges with only $37$ and $27$ samples for Kriging and PC-Kriging, respectively.
\begin{table}[!ht]
\caption{Four-branch function ($P=6$) performance comparison from \citet{Guo2021}.}
\label{tab:Ex1:sys}
    \centering
    \begin{tabular}{l c c c c}
    \hline 
         Method & $P_f$ & $\beta$ & $\varepsilon_{\beta}$ &  Total $N_{\text{eval}}$\\
         \hline 
         SuS (reference) & $4.484 \cdot 10^{-3}$ & $2.613$ &  $--$ & $~\sim 3 \cdot 10^{5}$ \\
         AK-SYS & $4.450 \cdot 10^{-3}$ & $2.616$  &  $1.034 \cdot 10^{-3}$ &  $63.3$ \\
         ALK-TCR & $4.480 \cdot 10^{-3}$ &  $2.614$ & $2.692 \cdot 10^{-4}$ &  $61.3$ \\
         ALK-SIS & $4.450 \cdot 10^{-3}$ & $2.616$ &  $1.034 \cdot 10^{-4}$ &  $62.4$ \\
         Proposed (KRG) & $ 4.500 \cdot 10^{-3}$ & $2.616$ & $ 8.496 \cdot 10^{-4}$ & $37$ \\
         Proposed (PCK) & $4.500 \cdot 10^{-3}$ & $2.616$ & $ 8.652 \cdot 10^{-4}$ & $27$ \\
         \hline
    \end{tabular}
\end{table}

\subsection{Roof truss problem}\label{sec:Example2}
\begin{figure}[!ht]
	\centering
	\includegraphics[width=0.5\textwidth]{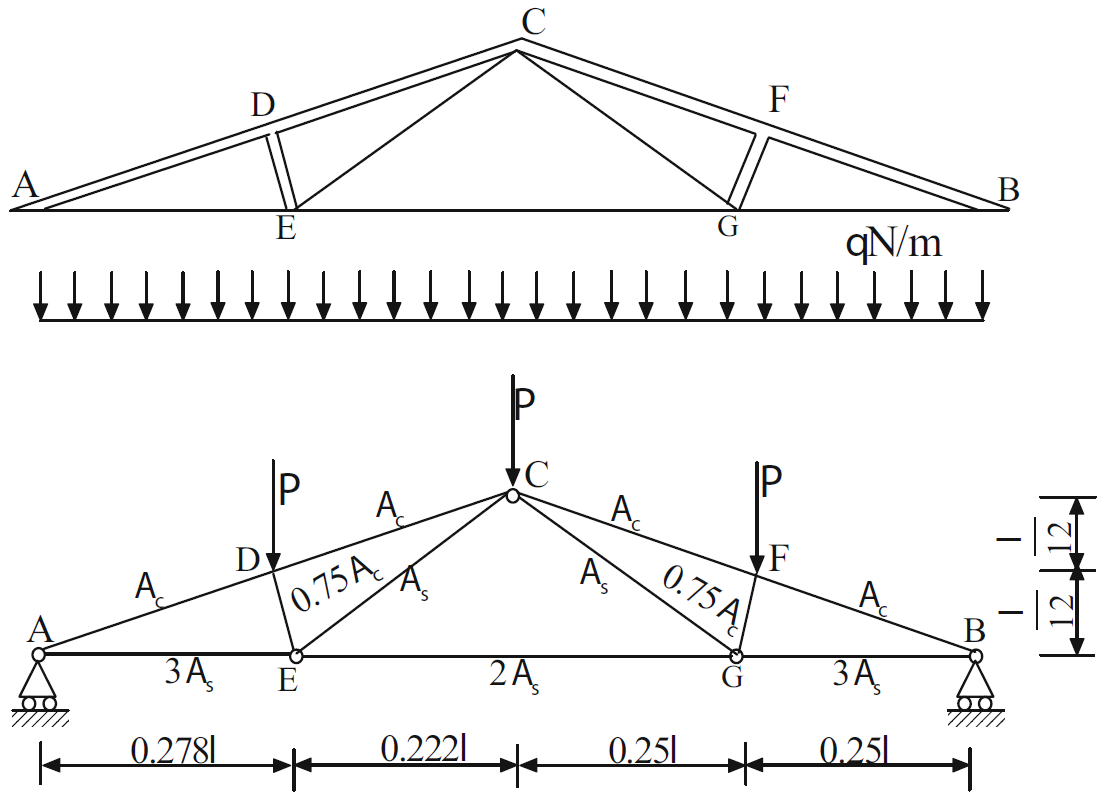}
 \caption{Example 2 - Schematic diagram of the roof structure (from \citet{Yun2019}).}
	\label{fig:roof}
\end{figure}

In this example, we consider a roof truss structure which consists of an assembly of beams and bars made of concrete and steel, as illustrated in Figure~\ref{fig:roof}. 
The structure is subjected to a distributed load on the top beams, which is equivalently modelled as point loads of magnitude $q l /4$ on nodes D and F and $ql/2$ on C, where $q$ is a line load and $l$ is the length of the roof base. It presents three distinct failure modes. The first one is related to the displacement of node C at the tip of the roof, which is expected to be below a critical threshold set at $3$ cm. It is represented by the limit state $g_1$ in Eq.~\eqref{eq:g_roof}, where $A_c$ and $E_c$ (resp. $A_s$ and $E_s$) are the cross-sectional areas of the concrete (resp. steel) bars and their constitutive materials Young's moduli. Failure in the second mode ($g_2$) occurs when the internal force in the bar AD, calculated as $1.185 \, q l$, is larger than its ultimate stress $f_c A_c$, where $f_c$ is the compressive strength of the bar AD. Similarly, failure from the third mode ($g_3$) occurs when the internal force in the bar EC, calculated as $0.75 \, ql$, is larger than its ultimate stress $f_s E_s$, where $f_s$ is the tensile strength of the bar EC. System failure is assumed when any of the limit states is violated, leading to a series system $h\prt{g_1\prt{\ve{x}},g_2\prt{\ve{x}},g_3\prt{\ve{x}}} = \min\prt{g_1\prt{\ve{x}},g_2\prt{\ve{x}},g_3\prt{\ve{x}}}$ whose components read
\begin{equation}
\label{eq:g_roof}
\begin{cases}
    g_1\prt{\ve{X}} = 0.03 - \frac{q l^2}{2} \prt{\frac{3.81}{A_c E_c} + \frac{1.13}{A_s E_s}}, \\
    g_2\prt{\ve{X}} = f_c A_c - 1.185 \, q l, \\
    g_3\prt{\ve{X}} = f_s A_s - 0.75 \, q l.
    \end{cases}
\end{equation}
The reference probability of failure is $3.417 \cdot 10^{-3}$, which corresponds to a reliability index of $2.705$ and is obtained from a subset simulation whose setting lead to a coefficient of variation of $0.2\%$. Note that for this specific example, if we were to assume that this roof is analysed using finite element modelling, a single run would yield all the three limit states. However, for the sake of comparison with other methods, we assume that each limit state is evaluated independently from the others.

The $8$ input parameters describing the system are gathered in Table~\ref{tab:Ex2} together with their probabilistic distributions. Only a subset of the random variables affects each limit state function, more precisely $6$ for $g_1$ and $4$ for $g_2$ and $g_3$. The initial EDs are therefore of sizes $N_1^{(0)} = 13$ and $N_2^{(0)} = N_3^{(0)} = 9$.
\begin{table}[!ht]
\caption{Probabilistic distribution of the random variables associated to the roof truss problem.}
\label{tab:Ex2}
    \centering
    \begin{tabular}{l c c c}
    \hline 
    Parameter & Distribution & Mean & Coef. of variation \\ \hline
    Uniform load $q$ (N/m) & Lognormal &  $20,000$ & $0.07$ \\
    Length $l$ (m) & Lognormal & $12$ &  $0.01$ \\
    Cross-sectional area $A_s$ (m$^2$) & Lognormal &  $9.82 \cdot 10^{-4}$ & $0.06$  \\
    Cross-sectional area $A_c$ (m$^2$) & Lognormal & $0.04$ &  $0.12$ \\
   Young's modulus $E_s$ (N/m$^2$) & Lognormal & $ 2 \cdot 10^{11}$ & $0.06$ \\
    Young's modulus $E_c$ (N/m$^2$) & Lognormal & $ 3 \cdot 10^{11}$ & $0.06$ \\
    Tensile strength $f_s$ (N/m$^2$) & Lognormal & $ 3.35 \cdot 10^{8}$ & $0.12$ \\
    Compressive strength $f_c$ (N/m$^2$) & Lognormal & $ 1.34 \cdot 10^{7}$ & $0.18$ \\
         \hline
    \end{tabular}
\end{table}

\begin{figure}[!ht]
	\centering
	\subfloat[Reliability index]{\label{fig:Ex2:Results_a}\includegraphics[width=0.45\textwidth]{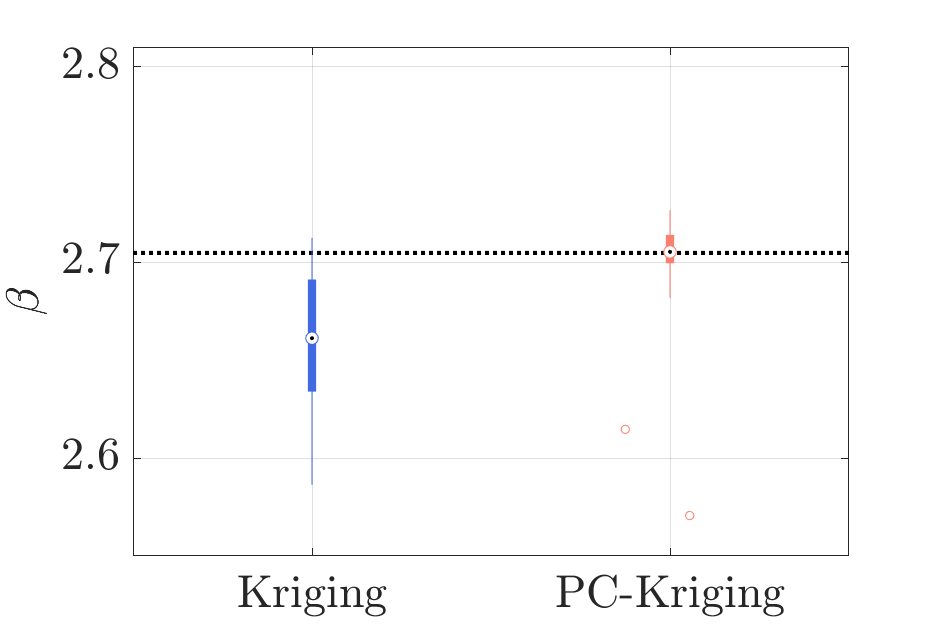}}%
	\subfloat[Number of model evaluations]{\label{fig:Ex2:Results_b}\includegraphics[width=0.45\textwidth]{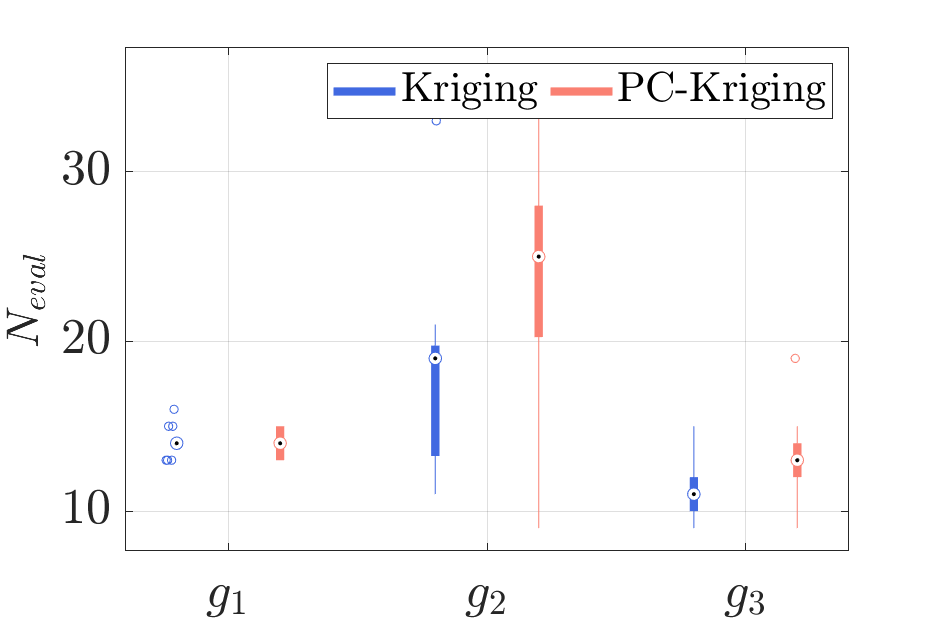}}%
	\caption{Example 2 - Boxplots summarizing the results of the $15$ repetitions.}
	\label{fig:Ex2:Results}
\end{figure}

Figure~\ref{fig:Ex2:Results_a} shows boxplots of the reliability index obtained by repeating the analysis $15$~times using Kriging and PC-Kriging. 
Both cases converge to the reference solution shown by the black dotted line, but with different degrees of accuracy.
While PC-Kriging yields extremely accurate and robust solutions, Kriging is slightly biased. 
This is due to premature convergence. 
This can be confirmed by the number of model evaluations shown in Figure~\ref{fig:Ex2:Results_b}, where Kriging requires significantly less model evaluations than PC-Kriging. 
In general, the limit-state $g_2$ is the most enriched, showing that it contributes the most to failure, and/or that it is non-linear in nature. 
Indeed, since it is of comparable complexity as $g_3$, which is evaluated much less, we may assume that $g_2$ has a larger contribution to failure. This can actually be confirmed by looking at the component failure probabilities. Using the latest surrogates in one of the repetitions and Monte Carlo simulation, these quantities can be estimated as $P_{f_2} = 3.3 \cdot 10^{-3}$ and $P_{f_3} = 3.5 \cdot 10^{-5}$. $P_{f_2}$ is indeed two order of magnitudes larger than $P_{f_3}$, which explains why $g_2$ is more often enriched than $g_3$ in the active learning scheme. 

We then compare these results against those gathered by \citet{Zhou2022} in Table~\ref{tab:Ex2:sys}. All results are accurate, with a relative error on the reliability index of the order of $10^{-4}$, except for Kriging for which the median relative error is in the order of $10^{-2}$.
In terms of model evaluations, our method is again more twice more efficient than the ones in the literature. 
PC-Kriging for instance, requires $51$ model evaluations in total, while this number grows to $114$, $120$ and $125$ for the methods in the literature. 
\begin{table}[!ht]
\caption{Roof structure results comparison with \citet{Zhou2022}.}
\label{tab:Ex2:sys}
    \centering
    \begin{tabular}{l c c c c c c c}
    \hline 
         Method & $P_f$ & $\beta$ & $\varepsilon_{\beta}$ & $N_{1}$ & $N_{2}$ & $N_{3}$ & $N_{\text{eval}}$\\
         \hline 
         SuS (reference) & $3.417 \cdot 10^{-3}$ & $2.705$ &  $--$ & & &  & $\sim 4 \cdot 10^5$ \\
         AK-SYS & $3.337 \cdot 10^{-3}$ & $2.713$  &  $3.036 \cdot 10^{-3}$ &  $32.1$ & $53.9$ & $34.9$ & $120.4$ \\
         ALK-TCR & $3.376 \cdot 10^{-3}$ & $2.709$  &  $1.557 \cdot 10^{-3}$ &  $32$ & $56$ & $37.1$ & $125.1$ \\
         IK-SRA & $3.350 \cdot 10^{-3}$ & $2.711$  &  $2.297 \cdot 10^{-3}$ &  $32$ & $48.8$ & $33.2$ & $114.0$ \\
         Proposed (KRG) & $3.893 \cdot 10^{-3}$ & $2.661$  &  $1.610 \cdot 10^{-2}$ &  $14$ & $19$ & $11$ & $43$ \\
         Proposed (PCK) & $3.412 \cdot 10^{-3}$ & $2.705$  &  $2.062 \cdot 10^{-4}$ &  $14$ & $25$ & $13$ & $51$ \\
         \hline
    \end{tabular}
\end{table}

\begin{figure}[!ht]
	\centering
	\subfloat[Reliability index]{\label{fig:Ex2:Results:eps0002_a}\includegraphics[width=0.45\textwidth]{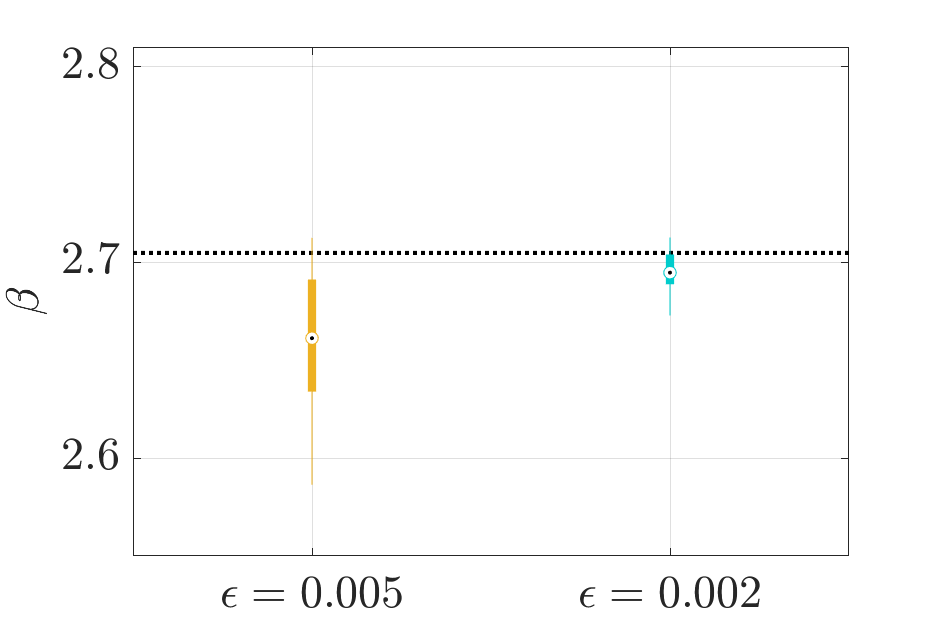}}%
	\subfloat[Number of model evaluations]{\label{fig:Ex2:Results:eps0002_b}\includegraphics[width=0.45\textwidth]{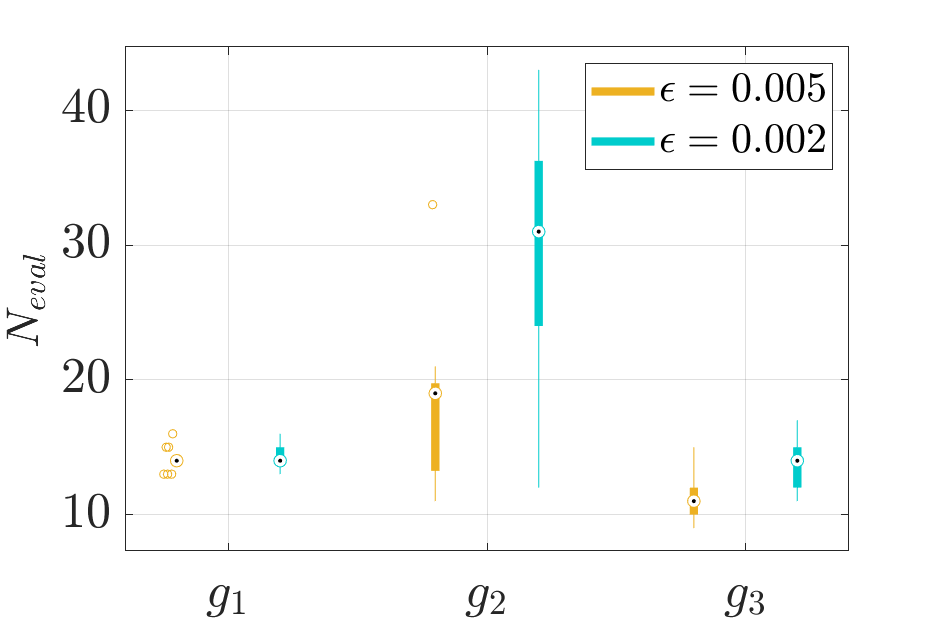}}%
	\caption{Example 2 - Results of the $15$ repetitions using Kriging with two different stopping criterion thresholds.}
	\label{fig:Ex2:Results:eps0002}
\end{figure}

Kriging requires even less model evaluations, but at the same time is slightly less accurate (although $1.6 \%$ error on the reliability index $\beta$ is more than enough in practice). 
For a fairer comparison, we run the Kriging-based approach, while setting the convergence criterion to $\bar{\varepsilon} = 0.002$ (previously set at $0.005$).
Figure~\ref{fig:Ex2:Results:eps0002} shows the resulting reliability index and model evaluations. As expected, they yield more accurate results. The median relative error on $\beta$ is decreased to $0.37 \%$, while the median number of model evaluations increases to $59$. Note that this remains much smaller than the cost of the benchmark methods shown in Table~\ref{tab:Ex1:sys}.
\subsection{Transmission tower}\label{sec:Ex3}
\begin{figure}[!ht]
    \centering\includegraphics[width=.9\textwidth,clip=true,trim=0 40 0 0]{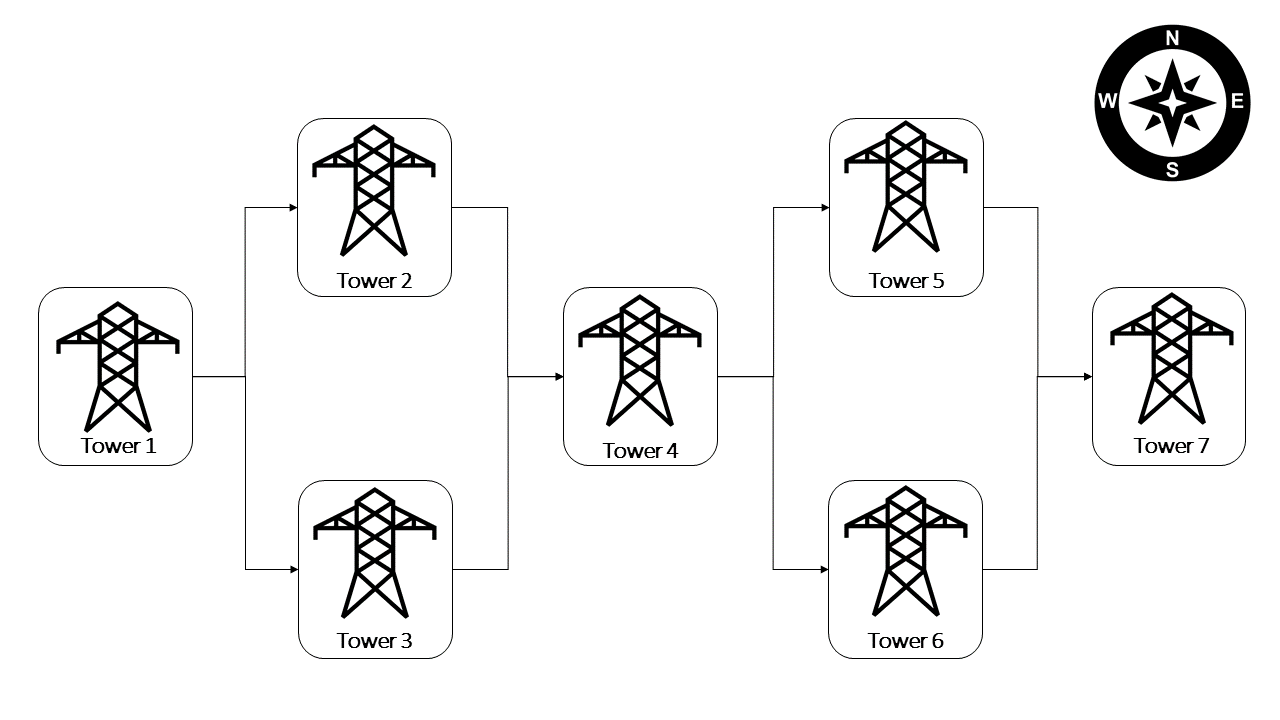}
    \caption{Example 3 - Schematic representation of the power transmission system.}
    \label{fig:Ex3:TowerSystem}
\end{figure}

In this final example, we validate the methodology on a complex engineering system related to power transmission. The system consists of $7$ towers, illustrated in Figure~\ref{fig:Ex3:TowerSystem}, and disposed such that electricity is transmitted from Tower 1 to Tower 7. Each tower is subjected to its own weight and to cable weights. Furthermore, they are in a field with extreme weather conditions and are therefore also subject to wind force.

\begin{figure}[!ht]
    \centering
    \includegraphics[width=0.35\textwidth]{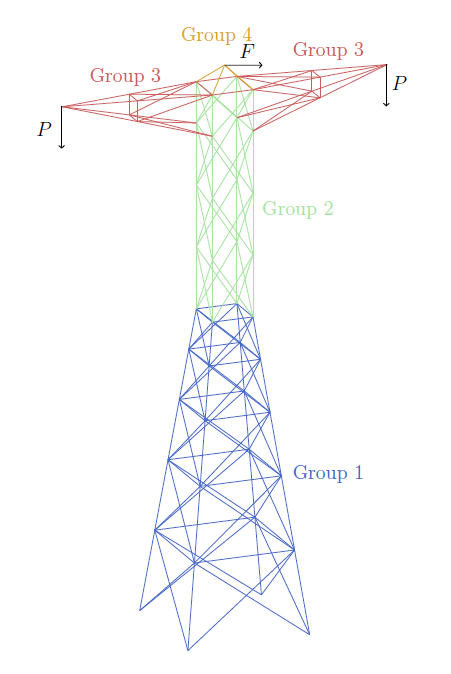}
    \caption{Example 3 - Finite element model of a single transmission tower.}
    \label{fig:Ex3:Tower}
\end{figure}

Figure~\ref{fig:Ex3:Tower} illustrates an archetype of the towers. They consist of $4$ groups of bars as shown by the different colours. 
Each group is characterized by its cross-sectional area and its constitutive material Young's modulus. 
We further consider that the towers within each row are similar (\emph{front row}: Towers $\#2$ and $\#5$, \emph{middle row}: Towers $\#1$, $\#4$  and $\#7$ and \emph{rear row}: Towers $\#3$  and $\#6$). 
Their corresponding parameters are denoted by $\acc{\ve{A}_f, \ve{E}_f}$, $\acc{\ve{A}_m, \ve{E}_m}$ and $\acc{\ve{A}_r, \ve{E}_r}$.
The cable weights $P$ are applied vertically on the extreme hands of the towers while the wind loads $F_1, \ldots, F_7$ are applied to the towers tips, as illustrated in Figure~\ref{fig:Ex3:Tower}. 
The direction of the wind for each tower is given by the angles $\alpha_1, \ldots, \alpha_7$.

To model the wind field, we assume that the wind comes from North to South and is therefore stronger on the front row. The dependency on the applied loads which is due to this wind field is described using copula theory \citep{Nelsen2006}. We consider two different pair copulas \citep{Bedford2002}:
\begin{itemize}
\item Wind force copula $C_{\textrm{WF}}$: The dependency among the wind forces $F_1$ to $F_7$ is modelled by a C-Vine copula with a truncation of second order. To account for the strong wind coming from North to South, we use a Gumbel copula in order to model upper tail dependence between towers on the front row (\ie Towers $\#2$ and $\#5$) and the remaining ones. It means that if we observe a strong wind force in the front row, we are likely to also observe strong ones in the remaining towers. We consider however that this correlation is strongly reduced past the first row, therefore we consider the remaining towers independent. In summary, the copula density is written as:
 \begin{equation}	c_{\textrm{WF}}\prt{\ve{u}} = \prod_{i = 1}^{6}\prod_{j=1}^{7-i}c_{j,j+i|\acc{1, \ldots, j-1}}\prt{u_{j|\acc{1, \ldots, j-1}},u_{j+i|\acc{1, \ldots, j-1}}},
\end{equation}
where the pair copulas are given by
\begin{equation}
	\begin{cases}
			c_{2,j}: \textrm{Gumbel}\prt{\theta = 5} \quad \textrm{for} \, j \neq 2 \\
			c_{5,j|2}: \textrm{Gumbel}\prt{\theta = 5} \quad \textrm{for} \, j \neq 2,5 \\
			c_{i,j|5,2}: \textrm{Independent}\quad \textrm{for} \, i \neq j, \, i,j \neq 2,5. 
	\end{cases}
\end{equation}
Using this setting the Person's (resp. Kendal's) correlation coefficients \citep{Kendall1970} between these seven parameters can be estimated to vary between $0.96$ and $0.99$ (resp. $0.80$ and $0.93$).
\item Wind angle copula $C_{\textrm{W}\alpha}$: The dependency between the wind angles $\alpha_1$ to $\alpha_7$ is modelled using a truncated C-Vine copula. We consider a $t$-copula to model both upper and lower tail dependencies. This means that if we observe a strong inclination (either east or west) in the front row, we are likely to observe a strong inclination in the remaining towers. The copula density is written as: 
\begin{equation}		c_{\textrm{W}\alpha}\prt{\ve{u}} = \prod_{i = 1}^{6}\prod_{j=1}^{7-i}c_{j,j+i|\acc{1, \ldots, j-1}}\prt{u_{j|\acc{1, \ldots, j-1}},u_{j+i|\acc{1, \ldots, j-1}}},
\end{equation}
where the pair copulas are given by
\begin{equation}
\begin{cases}
    c_{2,j}: t-\textrm{copula}\prt{\theta = 0.4, \nu = 2} \quad \textrm{for} \, j \neq 2 \\
    c_{5,j|2}: t-\textrm{copula}\prt{\theta = 0.4, \nu = 2} \quad \textrm{for} \, j \neq 2.5 \\
    c_{i,j|5,2}: \textrm{Independent}\quad \textrm{for} \, i \neq j, ij \neq 2,5.
\end{cases}
\end{equation}
\end{itemize}
Using this setting, the Person's (resp. Kendal's) correlation coefficient between these seven parameters can be estimated to vary between $0.23$ and $0.48$ (resp. $0.17$ and $0.36$). The reader is referred to \citet{Torre2019,UQdoc_20_102} for more details about the use of copula for modelling complex engineering systems and their implementation in \textsc{UQLab}.
	
The system is assumed to satisfy all operational requirements if electricity can be transmitted from Tower 1 to Tower 7. 
Failure for each tower is assumed when the maximum stress $S_{\textrm{max}_j}$ in any bar is larger than the yield stress $f_y$, which leads to the following limit states:
\begin{equation}
	\label{eq:gTower}
	g_j\prt{\ve{X_j}} = f_y - S_{\textrm{max}_j}, \qquad j = \acc{1, \ldots, 7}.
\end{equation}
The maximum stress is computed using a truss finite element code programmed in Matlab. 
For each tower, the limit state has an input $\ve{X}_j$ of dimension $12$ even though the entire set inputs $\ve{X}$ is of dimension $40$. 
Their marginal probabilistic distributions are given in Table~\ref{tab:Ex3:Param}.
\begin{table}[!ht]
\centering
\caption{Power transmission tower input marginals properties ($\theta_1$ and $\theta_2$ are the mean and the coefficient of variations for non-uniform variables, and the support bounds for uniform variables.)}
\begin{tabular}{ l  c  c  c  c }
\hline
	Group & Variable  & Distribution & $\theta_1$ & $\theta_2$ \\
	\hline
		\multirow{4}{4cm}{Cross-sectional area - front row} & $A_{f_1}$ (m$^2$)  & Gaussian & $1 \cdot 10^{-3}$  & $0.05$ \\
		& $A_{f_2}$ (m$^2$)  & Gaussian & $1 \cdot 10^{-3}$  & $0.05$ \\
		& $A_{f_3}$ (m$^2$) & Gaussian & $5 \cdot 10^{-3}$  & $0.05$ \\
		& $A_{f_4}$ (m$^2$)  & Gaussian & $5 \cdot 10^{-3}$  & $0.05$ \\ \cline{2-5}
		\multirow{4}{4cm}{Young's modulus - front row} & $E_{f_1}$ (Pa) & Gaussian & $210 \cdot 10^{9}$  & $0.05$ \\
		& $E_{f_2}$ (Pa) & Gaussian & $210 \cdot 10^{9}$  & $0.05$ \\
		& $E_{f_3}$ (Pa) & Gaussian & $210 \cdot 10^{9}$  & $0.05$ \\
		& $E_{f_4}$ (Pa)  & Gaussian & $210 \cdot 10^{9}$  & $0.05$ \\ \cline{2-5}
		\multirow{4}{4cm}{Cross-sectional area - midlle row} & $A_{m_1}$ (m$^2$)  & Gaussian & $1 \cdot 10^{-3}$  & $0.05$ \\
		& $A_{m_2}$ (m$^2$)  & Gaussian & $1 \cdot 10^{-3}$  & $0.05$ \\
		& $A_{m_3}$ (m$^2$) & Gaussian & $5 \cdot 10^{-3}$  & $0.05$ \\
		& $A_{m_4}$ (m$^2$)  & Gaussian & $5 \cdot 10^{-3}$  & $0.05$ \\ \cline{2-5}
		\multirow{4}{4cm}{Young's modulus - middle row} & $E_{m1}$ (Pa) & Gaussian & $210 \cdot 10^{9}$  & $0.05$ \\
		& $E_{m_2}$ (Pa) & Gaussian & $210 \cdot 10^{9}$  & $0.05$ \\
		& $E_{m_3}$ (Pa) & Gaussian & $210 \cdot 10^{9}$  & $0.05$ \\
		& $E_{m_4}$ (Pa)  & Gaussian & $210 \cdot 10^{9}$  & $0.05$ \\ \cline{2-5}
		\multirow{4}{4cm}{Cross-sectional area - rear row} & $A_{r_1}$ (m$^2$)  & Gaussian & $1 \cdot 10^{-3}$  & $0.05$ \\
		& $A_{r_2}$ (m$^2$)  & Gaussian & $1 \cdot 10^{-3}$  & $0.05$ \\
		& $A_{r_3}$ (m$^2$) & Gaussian & $5 \cdot 10^{-3}$  & $0.05$ \\
		& $A_{r_4}$ (m$^2$)  & Gaussian & $5 \cdot 10^{-3}$  & $0.05$ \\ \cline{2-5}
		\multirow{4}{4cm}{Young's modulus - rear row} & $E_{r_1}$ (Pa) & Gaussian & $210 \cdot 10^{9}$  & $0.05$ \\
		& $E_{r_2}$ (Pa) & Gaussian & $210 \cdot 10^{9}$  & $0.05$ \\
		& $E_{r_3}$ (Pa) & Gaussian & $210 \cdot 10^{9}$  & $0.05$ \\
		& $E_{r_4}$ (Pa)  & Gaussian & $210 \cdot 10^{9}$  & $0.05$ \\ \cline{2-5}
		Wind load - front row  & $F_2,F_5$ (N) & Gumbel & $3.25 \cdot 10^{4}$ & $0.30$  \\
		Wind load - middle row & $F_1,F_4,F_7$ (N) & Gumbel & $3.25 \cdot 10^{4}$ & $0.30$  \\
		Wind load - rear row & $F_3,F_6$ (N) & Gumbel & $3.25 \cdot 10^{4}$ & $0.30$  \\
		Wind angle & $\alpha_1,\dots,\alpha_7$ (deg) & Uniform & $-30$ & $30$ \\
		Cables weight & $P$ (N) & Gumbel & $1 \cdot 10^{4}$ & $0.30$  \\
		Yield stress & $f_y$ (MPa) & Lognormal & $355$ & $0.20$ \\
		\hline
\end{tabular}
\label{tab:Ex3:Param}
\end{table}

The reference solution, computed using subset simulation, is $P_f = 5.10 \cdot 10^{-4}$ ($\beta = 3.28$) with a coefficient of variation of $0.82\%$. This is obtained by using a large subset simulation which has required a total of $2.375 \cdot 10^6$ full evaluations of the computational model. 

For this example, we run the surrogate-based analysis with a stopping criterion set at $\bar{\varepsilon} = 10^{-3}$. However, we summarize the results for different intermediate thresholds in Table~\ref{tab:Ex3:Res}. The initial experimental design is of size $25$ for each limit state.
As with the previous examples, Kriging stops earlier than PC-Kriging but the latter is more accurate. In both cases though, the solutions are relatively accurate with an error in the order of $0.1\%$ to $1\%$ on the reliability index. The number of model evaluations is also remarkably small given the problem dimensionality and complexity. Note that this number is cumulated for all $7$ components.
\begin{table}[!ht]
\caption{Example 3 - Results using three different stopping criterion thresholds.}
\label{tab:Ex3:Res}
\centering
\begin{tabular}{l c c c c c}
		\hline 
		Method & $\bar{\varepsilon}$ & $\widehat{P}_f$ & $\widehat{\beta}$ & $\varepsilon_{\beta}$ &  Total $N_{\text{eval}}$\\
		\hline 
		SuS (reference) & $--$ & $5.102 \cdot 10^{-4}$ & $3.285$ &  $--$ & $2.375 \cdot 10^6$ \\ \cline{2-6}
		\multirow{3}{3cm}{Using Kriging} & $0.005$ & $4.435 \cdot 10^{-4}$  & $3.324$ & $1.195 \cdot 10^{-2}$ &  $245$ \\
		& $0.002$ & $4.542 \cdot 10^{-4}$  & $3.317$ & $9.921 \cdot 10^{-3}$ &  $281$ \\
		& $0.001$ & $4.542 \cdot 10^{-4}$  & $3.317$ & $9.921 \cdot 10^{-3}$ &  $281$ \\ \cline{2-6}
		\multirow{3}{3cm}{Using PC-Kriging} & $0.005$ & $5.396. \cdot 10^{-4}$  & $3.269$ & $4.827 \cdot 10^{-3}$ &  $221$ \\
		& $0.002$ & $5.420 \cdot 10^{-4}$  & $3.268$ & $5.198 \cdot 10^{-3}$ &  $248$ \\
		& $0.001$ & $5.390 \cdot 10^{-4}$  & $3.269$ & $4.720 \cdot 10^{-3}$ &  $327$ \\
		\hline
	\end{tabular}
\end{table}

\begin{figure}[!ht]
	\centering
	\subfloat[Kriging]{\label{fig:Ex3:Neval_a}\includegraphics[width=0.49\textwidth]{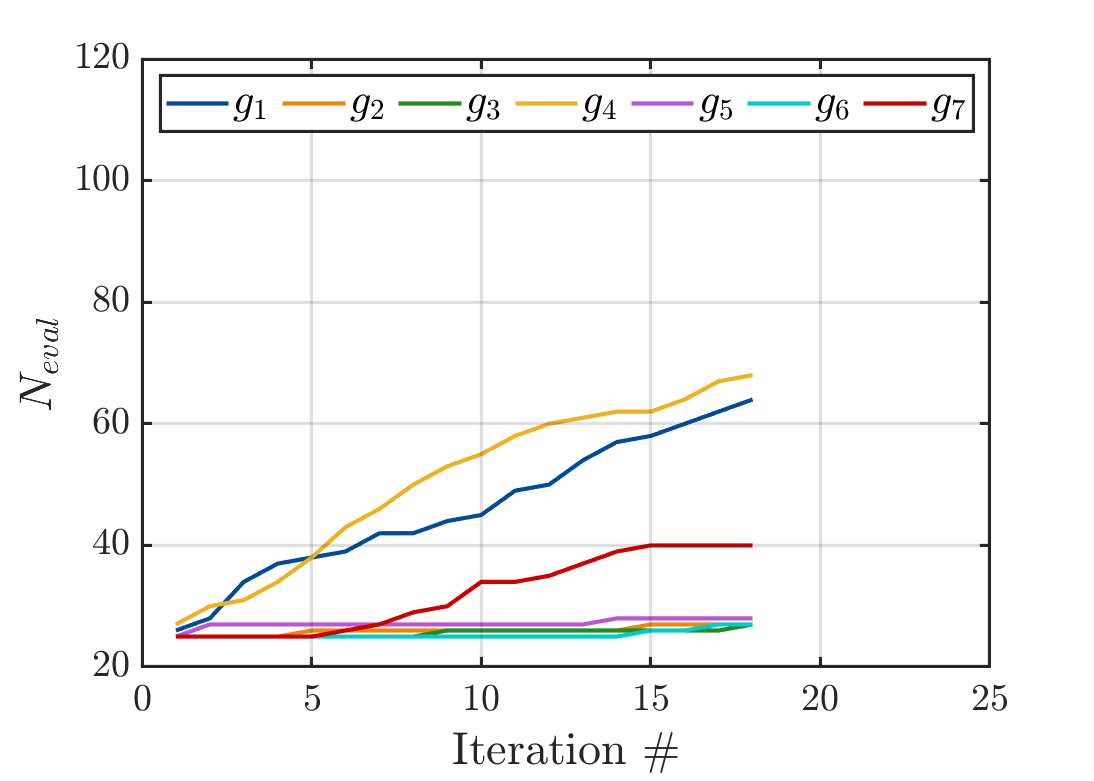}}%
	\subfloat[PC-Kriging]{\label{fig:Ex3:Neval_b}\includegraphics[width=0.49\textwidth]{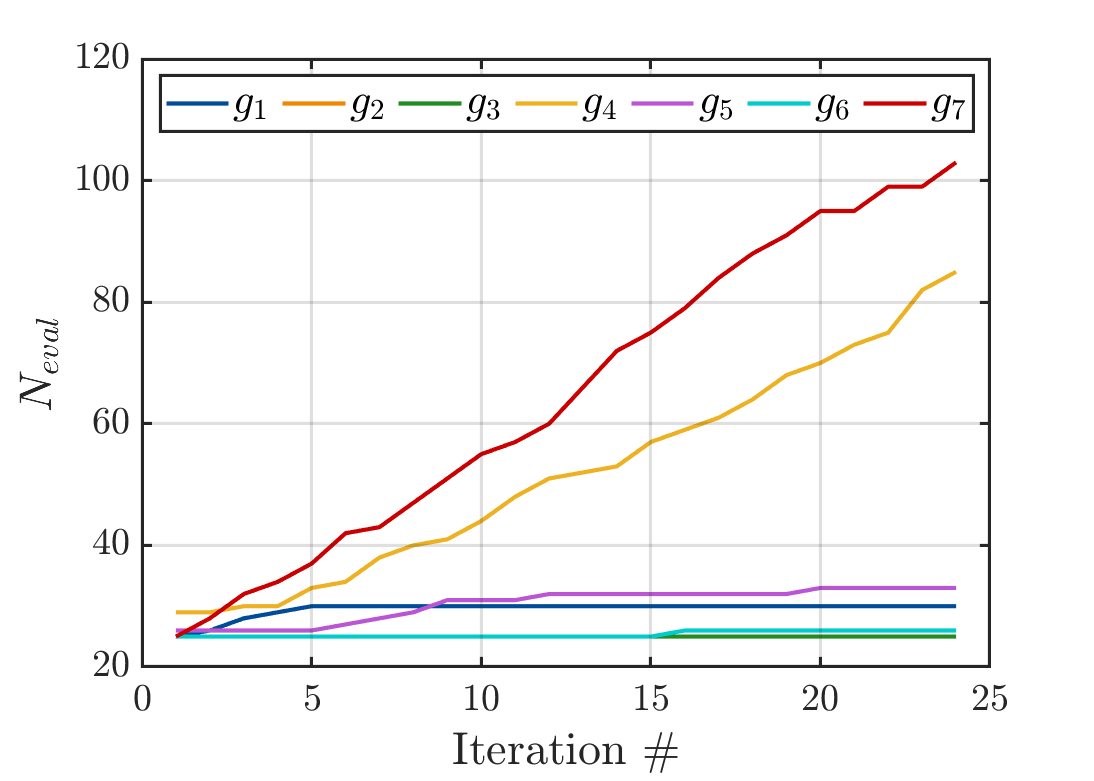}}%
	\caption{Example 3 - Evolution of the number of model evaluations for each component.}
	\label{fig:Ex3:Neval}
\end{figure}

Figure~\ref{fig:Ex3:Neval} shows the evolution of the number of model evaluations for each component throughout the enrichment iterations. Limit states $\#1$, $\#4$ and $\#7$, which are the most crucial, are the ones that are enriched the most. This is due to the fact that limits states $\#2$ and $\#5$ are in parallel with $\#3$ and $\#6$, hence making them less relevant to system failure.  

\begin{figure}[!ht]
\centering  \includegraphics[width=0.6\textwidth]{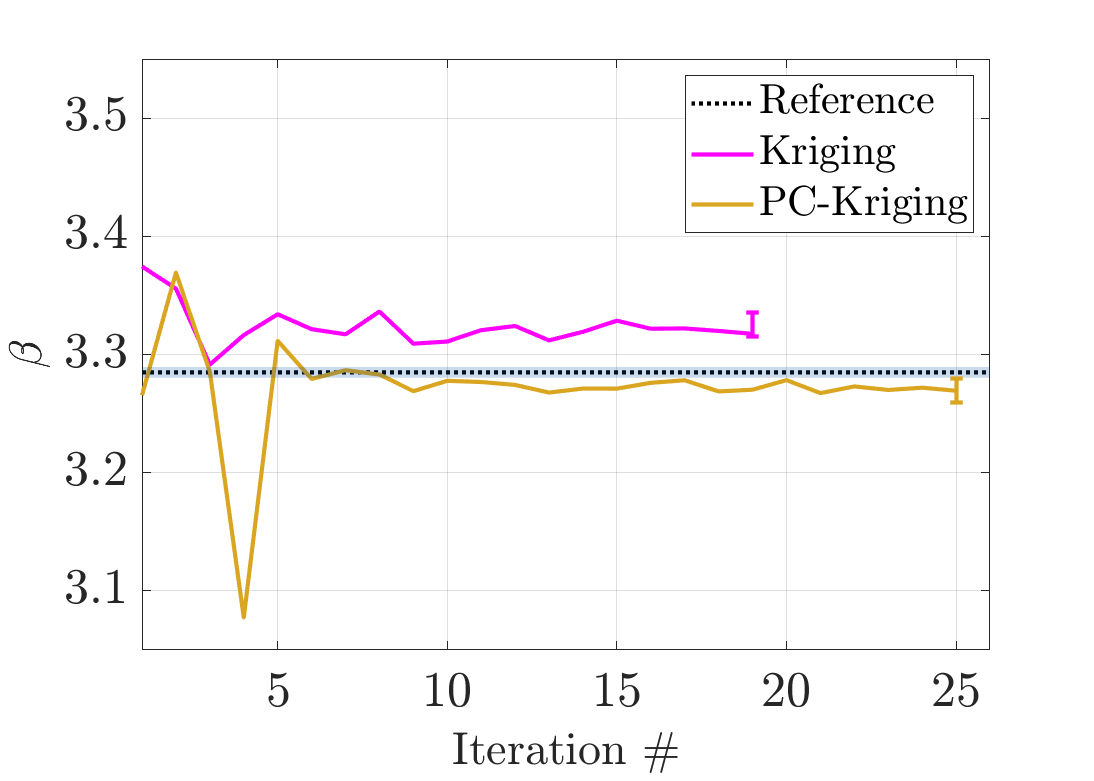}
\caption{Example 3 - Evolution of the reliability index.}
\label{fig:Ex3:Conv}
\end{figure}

Finally, Figure~\ref{fig:Ex3:Conv} shows the convergence of the reliability index. While limit state $g_7$ is enriched less than $g_1$ and $g_4$ for Kriging, $g_1$ is hardly enriched for PC-Kriging. These may explain the bias observed in the convergence curves. We can indeed again see that Kriging converges earlier but to a slightly biased solution. PC-Kriging yields a more accurate solution for which the $95 \%$ confidence interval is within that of the reference solution. These two cases however, together with the previous example, confirm the need to develop more robust stopping criteria for the proposed methodology. 

\section{Conclusion}
The literature on structural reliability analysis generally distinguishes component from system reliability analysis. While the former considers a single limit state function, the latter models more complex and realistic systems using multiple limit states. 
Component reliability has been largely studied and can be very efficiently solved using active learning approaches. Such approaches are, however, sub-optimal in the  context of system reliability because of the presence of multiple failure domains and the uneven contribution of each limit state to system failure.

In this work, we propose an active learning strategy that aims at efficiently solving system reliability problems in any arbitrary configuration. 
To this end, we consider approximating each limit state function by a separate surrogate model and devise an enrichment scheme that allows updating only the experimental design of the most influential limit states at each iteration. 
The efficiency of the proposed method stems from (i) a new learning function adapted to systems in arbitrary configuration, which does not assume any typical configuration (series or parallel), (ii) the use  of density-based clustering of the enrichment candidates to automatically identify the different failure modes and enrich them separately and (iii) the use of Sobol' sensitivity analysis to update only the limit states that effectively contribute to system failure.

The algorithm is validated on two analytical examples: a $2$-dimensional four-branch function and an $8$-dimensional roof structure. Both examples show that the proposed method is efficient and robust. 
The results are compared against similar methods in the literature and are shown to outperform them. On average the number of calls to the limit state functions is divided by two compared to the most recent state-of-the-art active learning methods.
Finally, a complex engineering system comprising 7 towers under extreme weather conditions is analyzed. This case corresponds to a general system (series or parallel systems) with a total of $40$ input random variables. 
In this case study, the developed method is  shown to converge to the reference solution using a very limited number of model evaluations (in the range of $250-300$), in particular when considering the problem dimensionality and complexity.

The proposed methodology inherits limitations of Kriging/PC-Kriging and DBSCAN in terms of dimensionality. While approximating each limit-state separately helps when only the system is of large dimension (as shown in the transmission tower example), the approach may break if some of the components are high-dimensional. For DBSCAN, we have proposed using dimensionality reduction. This may also be applied for building the surrogate models if the dimension of the component limit-state is too large for traditional Kriging/PC-Kriging. 

The methodology we propose focuses on the enrichment scheme. Nonetheless, other aspects may affect its efficiency, especially when dealing with complex failure domain topologies or extremely low probabilities of failure. In such scenarios, the robustness of the methodology can be enhanced by using a larger initial experimental design, or one that is more widely distributed. This approach increases the likelihood of obtaining an initial sample within or in close proximity to the failure domain, which in turn reduces the risk of overlooking a failure domain.

 

\appendix
 
\section{Density-based clustering (DBSCAN)}\label{app:dbscan}
DBSCAN is a hierarchical clustering algorithm proposed by \citet{Ester1996}. Its main advantages are that it can discover clusters of arbitrary shapes, does not require a pre-defined number of clusters to identify and is efficient on large datasets. 

DBSCAN is a density-based method that relies on the premise that a cluster is a subset of data points with high density separated to another cluster by an area of low point density. To formalize this, DBSCAN defines the $\varepsilon$-neighborhood of a point $\ve{x}$ as the set of points whose distance to $\ve{x}$ is smaller than $\varepsilon$, \emph{i.e.},
\begin{equation}
\mathcal{N}_\varepsilon\prt{\ve{x}} = \acc{\ve{s} \in \mathcal{X}: d\prt{\ve{x},\ve{s}} \leq \varepsilon},
\end{equation}
where $d\prt{\bullet,\bullet}$ is a distance measure. The algorithm is independent of the considered distance measure. We consider here the Euclidean distance. 

Using the concept of $\varepsilon$-neighborhood, the sample points can then be categorized into:
\begin{itemize}
\item \emph{Core points}, which are points that contain a minimum of $N_{\textrm{min}}$ points in their $\varepsilon$-neighborhood. They constitute the basis for defining a cluster, as each cluster must contain at least one core point. 
\item \emph{Border points}, which are points that have a core point but fewer than $N_{\textrm{min}}$ points in their $\varepsilon$-neighborhood.
\item \emph{Noise points}, which are neither core nor border points. They do not belong to any cluster.
\end{itemize}

DBSCAN uses these definitions to recursively partition the data set into clusters of arbitrary shapes. Schematically, the algorithm starts by randomly picking an unlabeled point in the data set. It then gathers points from its $\varepsilon$-neighborhood and if there are at least $N_{\textrm{min}}$ of them, then they all constitute a new cluster. The neighborhood is then recursively extended for each point in the current neighborhood until no point can be added. The algorithm restarts then with the next unlabeled point, which will then be part of a new cluster.

The algorithm can be summarized as follows:
\begin{algorithm}[!ht]
	\caption{DBSCAN \citep{Ester1996}}
	\begin{algorithmic}[1]
	\Require{}
	\Statex Define $\varepsilon$, the minimal distance to define the close neighborhood of a point
	\Statex Define $N_{\textrm{min}}$, the minimal number of points required to consider a region dense enough to be a cluster 
        \Statex Mark all points as unlabelled
        \Statex Define the unlabelled points as $\mathcal{X}_{\textrm{unlab}}$
	\Statex \hrulefill%
	\For{$\ve{x} \in \mathcal{X}_{\textrm{unlab}}$}
        \If {$\ve{x}$ {is unlabelled}}
        \State Mark $\ve{x}$ as labelled
        \State Find the set of points belonging to $\mathcal{N}_\varepsilon\prt{\ve{x}}$
        \If {$\abs{\mathcal{N}_\varepsilon\prt{\ve{x}}} < N_{\textrm{min}}$}
        \State Label $\ve{x}$ as noise point
        \Else
        \State Label $\ve{x}$ as a core point
        \State Start a new cluster and add $\ve{x}$ to this cluster
        \For{Each point $s$ in $\mathcal{N}_\varepsilon\prt{\ve{x}}$ }
        \State Assign $\ve{s}$ to current cluster
        \State Get the $\varepsilon$-Neighborhood of $s$: $\mathcal{N}_\varepsilon\prt{\ve{s}}$
        \If{$\abs{\mathcal{N}_\varepsilon\prt{\ve{s}}} \geq N_{\textrm{min}}$}
        $\mathcal{N}_\varepsilon\prt{\ve{x}} \leftarrow \mathcal{N}_\varepsilon\prt{\ve{x}} \cup \mathcal{N}_\varepsilon\prt{\ve{s}}$ \EndIf
        \EndFor
        \EndIf
        \EndIf
	\EndFor
	\end{algorithmic}
	\label{Alg:DBSCAN}
\end{algorithm}
\clearpage

\newpage
\bibliographystyle{chicago}
\bibliography{sysALR_RESS2023_Ref}

\end{document}